\documentclass[a4paper, english]{amsart}
\usepackage[utf8]{inputenc}

\usepackage{graphicx}
\usepackage{amsthm}
\usepackage{amsfonts}
\usepackage{mathtools}
\usepackage{amssymb}
\usepackage{paralist}
\usepackage[all,cmtip]{xy}
\usepackage{enumitem}
\usepackage{kantlipsum} 
\usepackage{makecell}

\usepackage[colorlinks=true, linkcolor=blue, citecolor=blue, linktocpage=true]{hyperref}
\usepackage[capitalise, nameinlink]{cleveref}
\crefname{subsection}{Subsection}{Subsections}
\crefname{equation}{}{}

\renewcommand\le\leqslant
\renewcommand\ge\geqslant

\calclayout

\usepackage[maxalphanames=4, minalphanames=4, maxbibnames=99,backend=bibtex, style=alphabetic]{biblatex} 
\AtBeginBibliography{\footnotesize}

\setlist[itemize]{align=parleft,left=0pt..12pt,topsep=0pt,label={\(\bullet\)}}
\setlist[enumerate]{align=parleft,left=0pt..18pt,topsep=0pt}

\newcommand{\fg}[0]{\mathfrak{g}}
\newcommand{\fn}[0]{\mathfrak{n}}

\newcommand{\fh}[0]{\mathfrak{h}}

\newcommand{\bC}[0]{\mathbb{C}}
\newcommand{\bN}[0]{\mathbb{N}}
\newcommand{\bZ}[0]{\mathbb{Z}}

\newcommand{\sheafO}[0]{\mathcal{O}}

\usepackage{amsthm}

\numberwithin{equation}{section}
\renewcommand*{\theequation}{%
  \ifnum\value{section}>0 %
    \thesection.%
  \fi
    \arabic{equation}%
}

\makeatletter
\def\th@plain{%
  \thm@notefont{}
  \itshape 
}
\def\th@definition{%
  \thm@notefont{}
  \normalfont 
}
\makeatother

\theoremstyle{plain}

\newtheoremstyle{example_style}
  {5pt} 
  {5pt} 
  {} 
  {} 
  {\itshape} 
  {.} 
  {10pt} 
  {} 

\theoremstyle{example_style}

\theoremstyle{definition}

\theoremstyle{example_style}

\title{The \(r\)-matrix structure of Hitchin systems via loop group uniformization}
\author{Raschid Abedin}\address{
ETH Z\"urich\\
Department of Mathematics \\
R\"amistrasse 101\\ 8006 Zurich\\
Switzerland
}
\email{raschid.abedin@math.ethz.ch}

\begin{document}

\maketitle

\begin{abstract}
    In this work, the description of the moduli space of principal \(G\)-bundles as double quotient of loop groups is used to construct an \'etale local \(r\)-matrix for the Hitchin integrable system.
\end{abstract}

\section{Introduction}
In \cite{hitchin_systems}, Hitchin introduced a remarkable family of integrable systems associated to the moduli space of \(G\)-bundles on Riemann surfaces for a complex semisimple Lie group \(G\). Many classical integrable models can be obtained from these systems as well as their generalizations introduced in \cite{markman_spectral_curves,bottacin_symplectic_geometry} and the quantization of these systems is related to conformal field theory and the geometric Langlands correspondence \cite{beilinson_drinfeld_quantization}. 

One of the fundamental methods in the theory of integrable systems is the \(r\)-matrix approach. If a classical mechanical system can be described by a Lax matrix \(L\) taking values in a Lie algebra \(\fg\)
and the Poisson bracket of \(L\) can be written in the form
\begin{equation}\label{eq:intro_first_russian}
    \{L\otimes L\} = [1 \otimes L,r] - [L\otimes 1,r^{21}]
\end{equation}
for some \(\fg \otimes \fg\)-valued map \(r\), the mechanical system in question admits a natural family of conserved quantities. This family turns out to be complete in many important situations, making the mechanical system integrable. The tensor \(r\) is then called \(r\)-matrix of the integrable system.

While Hitchin systems have been extensively studied, the \(r\)-matrix structure of these integrable models is comparatively not so well understood. Using the Schottky parameterization of \(G\)-bundles on Riemann surfaces, an \(r\)-matrix for Hitchin models was first constructed in \cite{enriquez}. Later in \cite{dolgushev}, the Tyurin parameterization of vector bundles was used to give another construction of an \(r\)-matrix for Hitchin systems in the case that \(G = GL_n(\bC)\).

In this work, we develop a different \(r\)-matrix approach for Hitchin systems using the loop group uniformization of the moduli space of \(G\)-bundles. Following Felder \cite{felder_kzb}, we construct a solution to a dynamical version of the classical Yang-Baxter equation \'etale locally on this moduli space of \(G\)-bundles. We prove that this solution is an \(r\)-matrix of the Hitchin system. Furthermore, we explain how an extension of this \(r\)-matrix can be viewed as a higher genus analog of the geometric solutions to the classical Yang-Baxter equation constructed in \cite{cherednik_definition_of_tau,burban_galinat}.

In \cite{felder_kzb}, the aforementioned \(r\)-matrix is used to express the KZB equation and, in the critical limit, the Hamiltonians of a generalized Gaudin-type model. We will see that the \(r\)-matrix approach developed here combined with the formulas from \cite{felder_kzb} leads to an ad hoc quantization of the considered Hitchin system. This can be viewed as an explicit aspect of the Beilinson-Drinfeld quantization of the Hitchin system \cite{beilinson_drinfeld_quantization}.

\subsection*{Results}
Let \(X\) be a Riemann surface of genus \(g > 1\) and \(G\) be a semisimple complex connected algebraic group. Moreover, let \(L^+G\) (resp.\ \(LG\), resp.\ \(L^-G\)) be the loop group of \(G\)-valued functions on the formal neighbourhood \(D\) of a finite subset \(S \coloneqq \{p_1,\dots,p_n\} \subseteq X\) (resp.\ on \(D^\circ \coloneqq D \setminus S\), resp.\ on \(X^\circ \coloneqq X\setminus S\)). We consider the moduli stack \(\underline{M}\) of \(G\)-bundles on \(X\). The loop group uniformization states that this stack can be naturally identified with the double quotient stack \(L^-G \setminus LG\,/\,L^+G\). The projection \(LG \to \underline{M}\) admits an \'etale quasi-section \(\sigma \colon U \to LG\) locally around regularly stable \(G\)-bundles.

Let \(L^\star \fg\) be the Lie algebra of \(L^\star G\) and \(K^\star \fg\) be the associated \(\fg\)-valued one-forms for \(\star \in \{\emptyset,+,-\}\). Fixing a non-degenerate invariant bilinear form on \(\fg\), we can define a canonical pairing \(B\) between \(L\fg\) and \(K\fg\). 

For every \(u \in U\) the subspace 
\[V(u) = \textnormal{Ad}(\sigma(u))L^-\fg \oplus \textnormal{Im}(\sigma(u)^{-1}d \sigma(u)) \subseteq L\fg\] 
is complementary to \(L^+\fg\). The projection onto \(L^+\fg\) associated to the decomposition 
\[L\fg = L^+\fg \oplus V(u)\] 
can be identified with a tensor series \(r(u) \in V(u)^\bot\widehat{\otimes} L^+\fg\) using the pairing \(B\). It turns out that the map \(u \mapsto r(u)\) is regular in an appropriate sense and \(r\) satisfies the following version of the dynamical classical Yang-Baxter equation:
\begin{equation}\label{eq:intro_dynamicalCYBE}
    \begin{split}&[r^{(13)},r^{(21)}] + [r^{(12)},r^{(23)}] + [r^{(13)},r^{(23)}] = \sum_{\alpha = 1}^m \left(\omega_\alpha^{(1)}\partial_\alpha r^{(23)} - \omega_\alpha^{(2)}\partial_\alpha r^{(13)}\right);
    \end{split}
\end{equation}
see Theorem \ref{thm:dynamicalCYBE}. Here, \(\{(u_\alpha,\partial_\alpha)\}_{\alpha = 1}^m\) is a coordinate system of \(TU\) and for every \(\alpha \in \{1,\dots,m\}\) the 1-form \(du_\alpha\) can be understood as a morphism \(\omega_\alpha \colon U \to K\fg\). The construction of \(r\) and Equation \eqref{eq:intro_dynamicalCYBE} are algebro-geometric reformulations of  \cite[Section 4.1 and Theorem 4.5]{felder_kzb} respectively. 

Consider the Hitchin integrable model with punctures at \(S\), i.e.\ the Hitchin system twisted by the line bundle \(\sheafO_X(S)\) in the sense of \cite{markman_spectral_curves,bottacin_symplectic_geometry}, where the finite subset \(S\subseteq X\) is considered as an effective divisor. This is the Hamiltonian system with phase space \(T^*M_{0,S}\), where \(M_{0,S}\) is the moduli space of regularly stable \(G\)-bundles on \(X\) with trivializations at the points of \(S\), and Hamiltonian given by \(H = \phi(L)\) for a canonical fiber-wise embedding \(L \colon T^*M_{0,S} \to K\fg\) and any \(G\)-invariant polynomial \(\phi\) on \(\fg^*\). After trivializing with \(\sigma\colon U \to LG\), we deduce that \(L\colon T^*M_{0,S}|_U \to K\fg\) defines a genuine Lax representation of this system, i.e.\
\begin{equation}\label{eq:intro_Lax}
    \frac{dL}{dt} = \{H,L\} = [Q,L]
\end{equation}
for an appropriate \(Q \colon T^*M_{0,S}|_U \to L\fg\); see Proposition \ref{prop:lax_rep}. 
The main result of this work is that \(r\) is an \(r\)-matrix of the punctured Hitchin system, i.e.\
\begin{equation}\label{eq:intro_rmatrix_of_Hitchin}
    \{L\otimes L\} = [1 \otimes L,r] - [L\otimes 1,r^{21}]
\end{equation}
holds; see Theorem \ref{thm:rmatrix_of_Hitchin} and Corollary \ref{cor:rmatrix_of_Hitchin}. We express a family of quadratic Hamiltonians of the punctured Hitchin system in terms of \(r\) and conclude that the quantization of this system was constructed in \cite{felder_kzb}; see Section \ref{sec:quantization}.

For \(t = \sum_{i = 1}^m \omega_\alpha \otimes \sigma^{-1}\partial_\alpha \sigma\), it turns out that 
\begin{equation}
    \rho(u) \coloneqq r(u) + t(u) \in \Gamma(X \times X^\circ,(\textnormal{Ad}(P(u)) \otimes \Omega_X) \boxtimes \textnormal{Ad}(P(u)))(\Delta))    
\end{equation}
has the identity element of 
\begin{equation}
    \Gamma(X^\circ,\textnormal{Ad}(P(u))\otimes \textnormal{Ad}(P(u))) \cong \textnormal{End}_{\sheafO_{X^\circ}}(\textnormal{Ad}(P(u))|_{X^\circ})
\end{equation}
as diagonal residue. Here, \(u \in U\), \(P(u)\) is the \(G\)-bundle associated to \(\sigma(u)\), and \(\Delta \subseteq X \times X\) is the diagonal divisor.
In particular, the extended \(r\)-matrix \(\rho = r + t\) is a point-wise Szeg\"o kernel (in the sense of \cite{benzvi_biswas_szegokernels}) and a higher genus analog of the geometric solutions to the generalized classical Yang-Baxter equation from \cite{cherednik_definition_of_tau,burban_galinat}.

The analogs of \eqref{eq:intro_dynamicalCYBE} and \eqref{eq:intro_rmatrix_of_Hitchin} for \(\rho\) are
\begin{equation}\label{thm:intro_extended_DCYBE}
    \begin{split}[\rho^{(13)},\rho^{(21)}] + [\rho^{(12)},\rho^{(23)}] + [\rho^{(13)},\rho^{(23)}]  = \sum_{\alpha = 1}^m \left(\omega_\alpha^{(1)}\nabla_\alpha \rho^{(23)} - \omega_\alpha^{(2)}\nabla_\alpha \rho^{(13)}\right),
    \end{split}
\end{equation}
where \(\nabla_\alpha = \partial_\alpha + \textnormal{ad}(\sigma^{-1}\partial_\alpha \sigma)\), and 
\begin{equation}\label{eq:intro_russian_rho}.
    [L\otimes L] = [1 \otimes L,\rho] - [L\otimes 1,\rho^{21}]
\end{equation}
respectively; see theorems \ref{thm:extended_DCYBE} \& \ref{thm:extended_rmatrix_of_hitchin}.

\subsection*{Structure}
In Section \ref{ch:loop_groups}, we discuss the basic properties of loop groups and loop algebras, their connection to moduli spaces, and the Hitchin systems. The construction of the \(r\)-matrix \(r\) as well as its relation to Hitchin systems and the quantization of these systems can be found in Section \ref{ch:rmatrices}. We conclude this work in Section \ref{ch:ex_rmatrices} with an examination of the extended \(r\)-matrix \(\rho = r + t\). In Appendix \ref{sec:not}, we give a brief overview of our notation.

\subsection*{Acknowledgments} 
I am thankful to Giovanni Felder for pointing out his work \cite{felder_kzb} to me and for productive discussions surrounding its content, which laid the foundations for this article. This work was supported by the DFG grant AB 940/1--1. It was also supported as a part of the NCCR SwissMAP, a National
Centre of Competence in Research, funded by the Swiss
National Science Foundation (grant number 205607).

\section{Loop groups, the moduli space of \(G\)-bundles, and the Hitchin system}\label{ch:loop_groups}

\subsection{The loop groups}
Let \(G \subseteq GL_n(\bC)\) be a semisimple connected complex algebraic group of dimension \(d\) defined by an ideal 
\begin{equation}\label{eq:ideal_of_G}
    I \subseteq \Gamma(GL_n(\bC),\sheafO_{GL_n(\bC)}) = \bC[(a_{ij})_{i,j = 1}^n,\textnormal{det}^{-1}].    
\end{equation}
Furthermore, let \(X\) be a complex irreducible smooth projective curve of genus \(g > 1\). Fix a finite subset \(S = \{p_1,\dots,p_\ell\}\subseteq X\) and write \(\mathfrak{m} = \mathfrak{m}_1 \cdot \ldots \cdot \mathfrak{m}_\ell \subseteq \sheafO_{X,S}\), where \(\mathfrak{m}_i\) is the maximal ideal of \(\sheafO_{X,p_i}\). The complement \(X^\circ \coloneqq X \setminus S\) of \(S\) is a smooth affine algebraic curve. 

\subsubsection{The complete loop group}\label{sec:complete_loop_group}
Let us write 
\begin{equation}\label{eq:O+}
    O^+ \coloneqq \widehat\sheafO_{X,S} = \varprojlim_k \sheafO_{X,S}/\mathfrak{m}^k = \prod_{i = 1}^\ell O_i^+\textnormal{, where } O_i^+\coloneqq \varprojlim_k\sheafO_{X,p_i}/\mathfrak{m}_i^k.
\end{equation}
for the completion of \(\sheafO_{X,S}\) at \(S\). Furthermore, let
\begin{equation}\label{eq:O}
    O \coloneqq \prod_{i = 1}^\ell O_i\textnormal{, where } O_i\coloneqq  (O_i^+\setminus\{0\})^{-1}O_i^+
\end{equation}
be the complete quotient ring of \(O^+\).

Consider the algebraic space \(LG\) defined by
\begin{equation}\label{eq:LG_algebraic_space}
    LG(R) \coloneqq \textnormal{Hom}_{\bC\textnormal{-alg}}(\Gamma(G,\sheafO_G),R\widehat{\otimes} O),
\end{equation}
where \(R\) is any \(\bC\)-algebra and \(R\widehat{\otimes} O = \varprojlim(R \otimes (O/\mathfrak{m}^k))\). 
It is well-known that \(LG\) is represented by an ind-affine group scheme, which will be denoted by the same symbol. 

Indeed, consider the affine scheme 
\begin{equation}
    M^{(N)} \coloneqq \mathfrak{m}^{-N}\textnormal{Mat}_{n \times n}(O^+) \cong \prod_{i = 1}^\ell \prod_{k = -N}^\infty \textnormal{Mat}_{n \times n}(\bC)    
\end{equation}
of infinite type and the affine subscheme \(L^{(N)}GL_n \subseteq M^{(N)}\) of invertible matrices \(A \in M^{(N)}\) such that \(A^{-1} \in M^{(N)}\). This subscheme can be identified with the affine subscheme of \(M^{(N)} \times M^{(N)}\) consisting of pairs \((A,B)\) such that \(AB = 1\). 
The group
\(LGL_n \coloneqq GL_n(O)\) gets its ind-affine structure from \(LGL_n = \bigcup_{N = 0}^\infty L^{(N)}GL_n\) and
\begin{equation}\label{eq:LG}
    LG = \{A \in GL_n(O)\mid p(A) = 0 \textnormal{ for all }p \in I\}
\end{equation}
is an ind-affine subgroup,
where \(I\) is the defining ideal of \(G\); see \eqref{eq:ideal_of_G}. In particular, \(LG\) obtains its ind-affine scheme structure from the filtration \(LG = \bigcup_{N = 0}^\infty L^{(N)}G\) for the affine subschemes \(L^{(N)}G = LG \cap L^{(N)}GL_n\) of \(L^{(N)}GL_n\).

\subsubsection{The inner loop group}\label{sec:inner_loop_group}
The subspace \(L^+G \coloneqq L^{(0)}G \subseteq LG\) is an affine group scheme of infinite type and is called inner loop group. It represents the algebraic space defined by 
\begin{equation}\label{eq:LG+_algebraic_space}
    L^+G(R) \coloneqq \textnormal{Hom}_{\bC\textnormal{-alg}}(\Gamma(G,\sheafO_G),R\widehat{\otimes} O^+)
\end{equation}
at any \(\bC\)-algebra \(R\).

\subsubsection{Coordinate representation}\label{sec:coordinates}
To make the above constructions more explicit, one can chose local coordinates
\(z_i\) of \(p_i\). Then \(\mathfrak{m}_i = (z_i)\), \(O_i^+ = \bC[\![z_i]\!]\), and \(O_i = \bC(\!(z_i)\!)\). Furthermore, 
\begin{equation}
    z = (z_1,\dots,z_\ell) \in O = \prod_{i = 1}^\ell O_i
\end{equation}
is a local coordinate of \(D\) and we can consider \(z_i \in O\) via \(O_i \subseteq O\).
After such a choice, we have \(R \widehat{\otimes} O^+ = \prod_{i = 1}^\ell R[\![z_i]\!]\) and \(R \widehat{\otimes} O = \prod_{i = 1}^\ell R(\!(z_i)\!)\) for every \(\bC\)-algebra \(R\).

\subsubsection{The outer loop group}\label{sec:outer_loop_group}
Consider
\begin{equation}\label{eq:O-}
    O^- \coloneqq \Gamma(X^\circ,\sheafO_X).
\end{equation}
The algebraic subspace of \(LG\) defined by 
\begin{equation}\label{eq:LG+_algebraic_space}
    R \longmapsto \textnormal{Hom}_{\bC\textnormal{-alg}}(\Gamma(G,\sheafO_G),R \otimes O^-),
\end{equation}
where \(R\) is any \(\bC\)-algebra, is represented by an ind-affine subgroup \(L^-G\subseteq LG\) called outer loop group of \(G\).

\subsubsection{The loop algebras}\label{sec:loop_algebras} The Lie algebra of \(L^\star G\) is \(L^\star\fg \coloneqq \fg \otimes O^*\) for \(\star \in  \{\emptyset,-,+\}\). More precisely, there is an Lie algebra isomorphism
\begin{equation}\label{eq:partial}
    \partial \colon L^\star\fg \to \textnormal{LDer}(\Gamma(L^\star G, \sheafO_{L^\star G})),    
\end{equation}
where \(\textnormal{LDer}(\Gamma(L^\star G, \sheafO_{L^\star G})\) denotes the left-invariant continuous derivations of \(\Gamma(L^\star G, \sheafO_{L^\star G})\) for \(\star \in  \{\emptyset,-,+\}\). Here, the topology on \(\Gamma(L^\star G, \sheafO_{L^\star G})\) is defined by the ind-structure of \(L^\star G\), i.e.\ its trivial for \(\star = +\).

We can think of this isomorphism as follows. Chose a triangular decomposition \(\fg = \fn_+ \oplus \fh \oplus \fn_-\) and observe that we have well-defined maps \(\exp \colon \fn_\pm\otimes O^\star \to L^\star G\) defined by \(a \mapsto \sum_{n = 0}^\infty \frac{a^n}{n!}\) (recall that \(G \subseteq \textnormal{GL}_n(\bC)\)) and we can write 
\begin{equation}\label{eq:partial_explicit}
    \partial_a \phi(g) = \frac{d}{ds} \phi(g\exp(as))\Big|_{s = 0}.    
\end{equation}
Since \(L^\star\fn_+ \oplus L^\star\fn_-\) generates \(L^\star\fg\) and \(\partial\) is a Lie algebra morphism, this defines \(\partial\) completely. 

Let us note that in coordinates (see Section \ref{sec:coordinates}), we simply have 
\begin{equation}
    L^+\fg = \prod_{i = 1}^\ell\fg[\![z_i]\!], L^-\fg = \fg \otimes O^- \subseteq L\fg = \prod_{i = 1}^\ell \fg(\!(z_i)\!).    
\end{equation}

\subsubsection{Geometric structure of loop algebras}\label{sec:geometry_of_loop_algebras}
Let \(\Omega_X\) be the sheaf of differential 1-forms on \(X\), write 
\begin{equation}\label{def:omega+-}
    \Omega^+ \coloneqq \widehat{\Omega}_{X,S},\textnormal{ and } \Omega^- \coloneqq \Gamma(X^\circ,\Omega_X).
\end{equation}
Furthermore, let \(\Omega \coloneqq O \Omega^+\) be the total quotient module of \(\Omega^+\). We have the usual residue map 
\begin{equation}\label{eq:def_res}
    \textnormal{res} \colon \Omega \longrightarrow \bC
\end{equation}
and we write
\begin{equation}\label{eq:K_notation}
    K^\star\fg \coloneqq \fg \otimes \Omega^\star \textnormal{ for }\star \in  \{\emptyset,-,+\}.
\end{equation}

Fixing a non-degenerate invariant bilinear form \(\kappa \colon \fg \times \fg\longrightarrow \bC\), we can define a pairing
\begin{equation}\label{eq:B}
    B \colon L\fg \times K\fg \stackrel{\kappa}\longrightarrow \Omega \stackrel{\textnormal{res}}\longrightarrow \bC.
\end{equation}
For every \(N \in \bZ\)
this pairing defines an isomorphism \((K\fg/K^{(-N)}\fg)^* \cong L^{(N)}\fg\) of vector spaces for \(L^{(N)}\fg = \mathfrak{m}^{-N}L^+\fg\) and \(K^{(-N)}\fg = \mathfrak{m}^{N}K^+\fg\). We can equip \(L^{(N)}\fg\) with the structure of an affine scheme of infinite type via
\begin{equation}\label{eq:scheme_structure_LNg}
    \Gamma(L^{(N)}\fg,\sheafO_{L^{(N)}\fg}) \coloneqq \textnormal{Sym}(K\fg/K^{(-N)}\fg).
\end{equation}
In particular, \(L^+\fg\) is an affine scheme of infinite type, while \(L\fg \coloneqq \bigcup_{N= 1}^\infty L^{(N)}\fg\) and \(L^-\fg \subseteq L\fg\) are ind-affine schemes.
In the same way, \(K^+\fg\) is an affine scheme and \(K\fg, K^-\fg\) are ind-affine schemes.

In coordinates (see Section \ref{sec:coordinates}), \(\Omega = \prod_{i = 1}^\ell \bC(\!(z_i)\!)dz_i\), \(K\fg = \prod_{i = 1}^\ell \fg(\!(z_i)\!)dz_i\), and 
\begin{equation}
    B\left(\sum_{i = 1}^\ell\sum_{k \in \bZ}a_{i,k}z_i^k,\sum_{i = 1}^\ell\sum_{k \in \bZ}b_{i,k}z_i^kdz_i\right) = \sum_{i = 1}^\ell \sum_{k \in \bZ} B(a_{i,k},b_{i,-k-1})
\end{equation}
for any \(a_{i,k},b_{i,k} \in \fg\).

\subsubsection{Vector fields on \(LG\)}\label{sec:vector_fields_on_LG} We have \(TLG \cong LG \times L\fg\) via left-trivialization. Therefore, sections of \(TLG\) (i.e.\ vector fields on \(LG\)) can be identified with morphisms \(LG \to L\fg\) and for two such maps \(a_1,a_2\) the commutator is given by 
\begin{equation}\label{eq:Lie_bracket_vector_fields_on_LG}
    [a_1,a_2](g) =  [a_1(g),a_2(g)] + \partial_{a_1(g)}a_2(g) - \partial_{a_2(g)}a_1(g).
\end{equation}
Here, \(\partial\) is the map \eqref{eq:partial}. In particular, left-invariant vector fields are precisely identified with constant functions, so elements of \(L\fg\).

\subsubsection{Poisson structure of \(T^*LG\)}\label{sec:poisson_structure_on_LG} Dual to Section \ref{sec:vector_fields_on_LG}, we have 
\begin{equation}\label{eq:trivialization_of_cotangent_LG}
    T^*LG \cong LG \times K\fg    
\end{equation}
where \(K\fg \cong \fg \otimes \Omega\); see \eqref{eq:K_notation}.
The product of ind-affine schemes is naturally an ind-affine scheme and we can write
\begin{equation}\label{eq:functions_on_cotangent_LG}
    \begin{split}
        \Gamma(T^*LG,\sheafO_{T^*LG})&= \Gamma(LG,\sheafO_{LG}) \widehat{\otimes} \Gamma(K\fg,\sheafO_{K\fg}) \\& \coloneqq \varprojlim_{N} \left(\Gamma(L^{(N)}G,\sheafO_{L^{(N)}G}) \otimes \textnormal{Sym}(L\fg/L^{(-N)}\fg)\right);
    \end{split}
\end{equation}
see e.g.\ \cite[Section 4]{kumar_kac_moody_groups}. 
The Poisson bracket of \(\Gamma(T^*LG,\sheafO_{T^*LG})\) is given uniquely as the continuous bi-derivative satisfying 
\begin{equation}
    \{f_1,f_2\} = 0, \{a_1,f_1\} = \partial_{a_1} f_1, \textnormal{ and }\{a_1,a_2\} = [a_1,a_2]
\end{equation}
for \(f_1,f_2 \in \Gamma(LG,\sheafO_{LG}), a,b \in L\fg \subseteq \Gamma(K\fg,\sheafO_{K\fg})\).

\subsection{Moduli spaces of \(G\)-bundles}
Let us briefly outline the connection between loop groups and the moduli space of \(G\)-bundles.

\subsubsection{The complete loop group \(LG\) as moduli space}\label{sec:LG_as_moduli}
Recall that a \(G\)-bundle \(P \to S\) over a scheme \(S\) is a scheme with \(G\)-action such that there exists an \'etale covering \(S' \to S\) admitting a \(G\)-equivariant isomorphism \(P \times_S S' \cong G \times S'\) of \(S'\)-schemes.

Let us write
\begin{equation}
    D \coloneqq \textnormal{Spec}(O^+) \textnormal{ and }D^\circ \coloneqq \textnormal{Spec}(O)
\end{equation}
for the formal neighbourhood of \(S\) and the punctured formal neighbourhood of \(S\) respectively.

To every \(\bC\)-algebra \(R\) and every \(g \in LG(R)\), one can associate a \(G\)-bundle \(P\) on \(X \times \textnormal{Spec}(R)\) by gluing the trivial bundles on \(D \times \textnormal{Spec}(R)\) and on \(X^\circ \times \textnormal{Spec}(R)\) together over \(D^\circ \times \textnormal{Spec}(R)\) using \(g\). This bundle comes with trivializations \(\varphi_+\) and \(\varphi_-\) on \(D \times \textnormal{Spec}(R)\) and \(X^\circ \times \textnormal{Spec}(R)\) respectively.

In more geometric terms, we can identify \(LG\) as algebraic space with the functor that maps \(R\) to triples \((P,\varphi_+,\varphi_-)\) over \(X \times \textnormal{Spec}(R)\) as above. In particular, if \(LG\) is considered as ind-affine scheme, a \(\bC\)-point \(g \in LG\) is identified with a triple \((P,\varphi_+,\varphi_-)\) of a \(G\)-bundle \(P\) on \(X\) with trivializations \(\varphi_+\) and \(\varphi_-\) on \(D\) and \(X^\circ\) respectively.

\subsubsection{The flag varieties as moduli space}\label{sec:flags} If we replace \(g\) with \(gh\) for \(h \in L^+G(R)\) in Section \ref{sec:LG_as_moduli}, the pair \((P,\varphi_-)\) is preserved while \(\varphi_+\) is changed. In this way, we can identify the stack of these pairs \((P,\varphi_-)\) with the quotient stack \(F^+ \coloneqq LG\,/\,L^+G\). This stack turns out to be an ind-projective scheme and is called the (inner) affine flag variety.  

Similarly, the pairs \((P,\varphi_+)\) can be identified with the quotient stack
\(F^-\coloneqq L^-G\setminus LG\). This stack turns out to be defined by a scheme of infinite type (see e.g. \cite[4.1.5. Proposition]{benzvi_frenkel_spectral_curves}) and we will call it the outer flag variety. In particular, a \(\bC\)-point \([g] \in F^-\) can be identified with a \(G\)-bundle \(P \to X\) equipped with a trivialization over \(D\).

\subsubsection{The moduli space of \(G\)-bundles and loop group uniformization} One can combine the constructions from the sections \ref{sec:LG_as_moduli} and \ref{sec:flags} in order to identify the moduli stack of \(G\)-bundles on \(X\) with the double quotient stack
\begin{equation}
    \underline{M} \coloneqq L^-G \setminus LG \,/\, L^+G. 
\end{equation}
This identification is also known as uniformization theorem. Let us note that \(\underline{M}\) turns out to be an honest stack, i.e.\ it cannot be represented by a scheme or ind-scheme.

However, the substack \(\underline{M}_0 \subseteq \underline{M}\) of \(G\)-bundles which are regularly stable, i.e.\ whose automorphism set coincides with the center of \(G\), turns out to be open and admits a coarse moduli space \(M_0\) which is a smooth quasi-projective variety.

\subsubsection{\(G\)-bundles with trivializations} Let \(\underline{M}_S\) be the stack of \(G\)-bundles \(Q\) with trivializations at the points in \(S\), i.e.\ which come equipped with \(G\)-equivariant isomorphisms \(Q|_{p_i} \cong G\) for \(i \in \{1,\dots,\ell\}\). Then, if we write \(L^{>0}G\) for the kernel of the evaluation \(L^+G \to G^\ell\) at the points in \(S\), we have
    \begin{equation}
        \underline{M}_S = L^-G\setminus LG\, /\, L^{>0}G = F^-/L^{>0}G,
    \end{equation}
Let us denote by \(M_{0,S}\) the coarse moduli space of regularly stable \(G\)-bundles on \(X\) with trivializations at the points in \(S\).

\subsection{Hitchin systems} Following \cite{beilinson_drinfeld_quantization}, the Poisson center of \(\overline{\textnormal{Sym}}(L\fg) = \Gamma(K\fg,\sheafO_{K\fg})\) is given by \(\overline{\textnormal{Sym}}(L\fg)^{L\fg} \subseteq \overline{\textnormal{Sym}}(L\fg)\) and
\begin{equation}
    \overline{\textnormal{Sym}}(L\fg) = \Gamma(K\fg,\sheafO_{K\fg}) \subseteq \Gamma(T^*LG,\sheafO_{T^*LG}) \cong \Gamma(LG,\sheafO_{LG}) \widehat{\otimes} \Gamma(K\fg,\sheafO_{K\fg})
\end{equation}
is an embedding of Poisson algebras. Therefore, \(\overline{\textnormal{Sym}}(L\fg)^{L\fg} \subseteq \Gamma(T^*LG,\sheafO_{T^*LG})\) is a set of Poisson commuting regular functions. 

The cotangent space \(T^*F^-\) of the outer flag \(F^-\) can be obtained by reduction from \(T^*LG\). Indeed, \(T^*F^- = \mu^{-1}(0)/L^-G\) for the moment map 
\begin{equation}
    \mu \colon T^*LG \cong LG \times K\fg \longrightarrow (L^-\fg)^* \cong K\fg/K^-\fg\,,\qquad (g,a) \longmapsto [-\textnormal{Ad}(g) a].
\end{equation}
In particular, the restriction map \(\sheafO_{T^*LG}\to \sheafO_{\mu^{-1}(0)}\) is a Poisson morphism and \(\sheafO_{T^*F^-} \subseteq \sheafO_{\mu^{-1}(0)}\) is the subsheaf of \(L^-G\)-periodic regular functions.

The subalgebra 
\begin{equation}
    \overline{\textnormal{Sym}}(L\fg)^{L\fg} \subseteq \Gamma(T^*LG,\sheafO_{T^*LG})
\end{equation} 
restricts to a subalgebra of Poisson commuting functions on \(\mu^{-1}(0)\) which are \(L^-G\)-invariant. Therefore, these define a set of Poisson commuting functions on \(T^*F^-\) as well. 

For any algebraic subgroups \(K \subseteq L^+G\) we can use another reduction in order to see that the cotangent space \(T^*(F^-/K)\) of the quotient stack \(F^-/K\) behaves nicely and, since the functions in \(\overline{\textnormal{Sym}}(L\fg)^{L\fg}\) are \(K\)-invariant, these define a set of commuting functions on \(T^*(F^-/K)\) as well. One can now consider the classical mechanical system with these Poisson-commuting functions as Hamiltonians. For appropriate \(K\), one obtains algebraically completely integrable systems in this way, which are called Hitchin systems. 

The original version, introduced by Hitchin in \cite{hitchin_systems} and transported to the stack language outlined here in \cite{beilinson_drinfeld_quantization}, is given by \(K = L^+G\). In this case \(F^-/K = \underline{M}\) is the moduli space of \(G\)-bundles. Another important case, which is the stack version of Hitchin systems discussed in e.g.\ \cite{markman_spectral_curves,bottacin_symplectic_geometry}, is that \(K\) is given by the kernel of the evaluation map \(L^+G \to G^\ell\) at the points in \(S\). Then \(F^-/K = \underline{M}_S\) is the moduli space of \(G\)-bundles with trivializations at \(S\). 

In this work, we consider the Hitchin system over \(\underline{M}_S\), or more precisely, its restriction over the coarse moduli space \(M_{0,S}\) of regularly stable \(G\)-bundles with trivializations at \(S\), and refer to it as punctured Hitchin system. We will see that it will be convenient to consider the infinite-dimensional classical mechanical system on \(T^*F^-\) first before passing to \(T^*M_{0,S}\) by reduction. Therefore, we refer to this analog of the Hitchin system as formal Hitchin system.

\subsubsection{Hitchin fibration}\label{sec:hitchin_fibration} Consider the affine scheme quotient \(C\coloneqq \textnormal{Spec}(\textnormal{Sym}(\fg)^G)\) of \(\fg^*\) by the coadjoint action. Using the natural \(\mathbb{G}_m\)-action on this scheme, we can consider its \(\Omega_X\)-twist \(C'\). The projection \(\fg^* \to C\) induces a map \(\textnormal{Ad}(P) \otimes \Omega_X \to C'\) for every \(G\)-bundle \(P\). Applying \(\Gamma(X^\circ,-)\) and using \(\Gamma(X^\circ,\textnormal{Ad}(P) \otimes \Omega_X) \cong T_{[g]}F^-\) for the \(G\)-bundle \(P\) defined by \([g] \in F^-\), this results in a fibration
\begin{equation}\label{eq:formal_hitchin_fibration}
    T^*F^- \to \Gamma(X^\circ,C) \eqqcolon Z    
\end{equation}
by varying \([g]\). 

The Hamiltonians coming from the Poisson center of \(K\fg\) generate the image of the corresponding morphism
\begin{equation}
    \Gamma(Z,\sheafO_Z) \longrightarrow \Gamma(T^*F^-,\sheafO_{T^*F^-}).
\end{equation} 
Here, we used that \(Z\) is a vector space with countable basis and hence has a natural structure of an ind-affine scheme isomorphic to \(\mathbb{A}^\infty_\bC\); see \cite[Examples 4.1.3.(3)]{kumar_kac_moody_groups}. 

There exist  \(\textnormal{rk}(\fg)\) homogeneous generators \(p_i \in \textnormal{Sym}(\fg)^G\) of degree \(d_i = e_i + 1\), where \(\{e_1,\dots,e_i\}\) are the exponents of \(\fg\). A choice of such generators provides an identification 
\[\textnormal{Sym}(\fg)^G \cong \bC[p_1,\dots,p_{\textnormal{rk}(\fg)}]\] 
and therefore an identification \(C' \cong \prod_{i = 1}^{\textnormal{rk}(\fg)}\Omega_X^{\otimes d_i}\). The fibration \eqref{eq:formal_hitchin_fibration} then takes the form
\begin{equation}\label{eq:formal_hitchin_fibration2}
    T^*F^- \to \bigoplus_{i = 1}^{\textnormal{rk}(\fg)}\Gamma(X^\circ,\Omega_X^{\otimes d_i}).
\end{equation}
Let us note that we always have \(d_1 = 2\) and the associated \(p_1\) can be chosen to be defined by the non-degenerate invariant bilinear form \(\kappa\) chosen in Section \ref{sec:loop_algebras}.

The fibration \eqref{eq:formal_hitchin_fibration2} can be adjusted to the Hitchin systems on \(\underline{M}, \underline{M}_S\) and \(M_{0,S}\). The resulting fibrations are called Hitchin fibrations. For the punctured Hitchin system on \(M_{0,S}\), which we will consider, the Hitchin fibration takes the form 
\begin{equation}
    T^*M_{0,S} \to \bigoplus_{i = 1}^{\textnormal{rk}(\fg)}\Gamma(X,\Omega_X(S)^{\otimes d_i}),    
\end{equation}
where \(S \subseteq X\) is considered as an effective divisor. 

\section{The \(r\)-matrix structure of the punctured Hitchin system}\label{ch:rmatrices}

\subsection{Local trivialization of \(F^-\)}\label{sec:trivialization_ofF-}\label{lem:trivialization_of_M0} Let \(F^\pm_0\) and \(LG_0\) be the preimage of the open substack \(\underline{M}_0 \subseteq \underline{M}\) of regularly stable \(G\)-bundles under the canonical projection \(F^\pm \to \underline{M}\) and \(LG \to \underline{M}\) respectively. 
The projection \(F^+_0 \to M_0\) admits local \'etale quasi-sections; see e.g.\ \cite{laszlo_hitchin_wzw}. Moreover, the projection \(LG \to F^+\) admits a section in the Zariski topology. Combined, we can see that \(LG_0 \to M_0\) admits a local \'etale quasi-section. 

This means that for every \(Q \in M_0\), where we recall that \(M_0\) is the coarse moduli space of regularly stable \(G\)-bundles, there exists an \'etale morphism \(P \colon U \to M_0\) and a morphism \(\sigma \colon U \to LG\) such that the diagram
\begin{equation}\label{eq:etale_quasisection}
    \xymatrix{U \ar[r]^\sigma\ar[d]_{P} & LG \\ M_0  & LG_0 \ar[l]\ar[u]_\subseteq},
\end{equation}
commutes and \(Q \in P(U)\). In other words, for every \(u \in U\), \(\sigma(u) \in LG\) defines the \(G\)-bundle \(P(u) \in M_0\). 

We fix the quasi-section \(\sigma \colon U \to LG\) and the \'etale open subset \(P \colon U \to M_0\) for the rest of this work. The effect of choosing a different pair \(\sigma, P\) on the following constructions will be outlined in Section \ref{sec:choosing_a_different_sigma}.
Let us note that the choice of this pair gives an identification \(F^-|_U \coloneqq U \times_{M_0} F^-_0 \cong U \times L^+G \) using
\begin{equation}
    U \times L^+G \longrightarrow F^-|_U\,,\qquad (u,g) \longmapsto [\sigma(u)]\cdot g,
\end{equation}
where \([\sigma(u)] \in F^- = L^-G \setminus LG\) is the equivalence class of \(\sigma(u) \in LG\). Similarly, \(M_{0,S}|_{U} = U \times_{M_0} M_{0,S} \cong U \times G^\ell\).

\subsubsection{Expressing sections using \(\sigma\) and \(P\)}\label{sec:sections_via_sigma}
Let us note that \(\sigma\) encodes important informations about the family of \(G\)-bundles defined by \(P \colon U \to M_0\). For instance,
\begin{equation}
    \Gamma(X^\circ,\textnormal{Ad}(P(u))) = \textnormal{Ad}(\sigma(u)^{-1})L^-\fg \textnormal{ and }\Gamma(X^\circ,\textnormal{Ad}(P(u)) \otimes \Omega_X) = \textnormal{Ad}(\sigma(u)^{-1})K^-\fg.
\end{equation}

\subsubsection{Expressing universal sections using \(\sigma\) and \(P\)}\label{sec:universal_sections_via_sigma} Let us write \(L^\star\fg(U)\) (resp.\ \(K^\star\fg(U)\)) for the regular maps \(U \to L^\star\fg\) (resp.\ \(U \to K^\star\fg\)), for \(\star \in \{\emptyset,+,-\}\). Moreover, let \(L_\sigma^-\fg(U)\) (resp.\ \(K_\sigma^-\fg(U)\)) denote the regular maps \(a\colon U \to L\fg\) (resp.\ \(a \colon U \to K\fg\)) such that \(\textnormal{Ad}(\sigma)a\) takes values in \(L^-\fg\) (resp.\ \(K^-\fg\)). 
An element in \(L^-_\sigma(U)\) is a family of sections over \(X^\circ\) of the adjoint \(G\)-bundles parametrized by \(\sigma\) that varies regularly over \(U\).

Using the identification \(F^-|_U \cong U \times L^+G\) we have
\begin{equation}
    \begin{split}
        &TF^-|_U \coloneqq T(F^-|_U) \cong TU \times L^+G \times L^+\fg;\\ 
        &T^*F^-|_U \coloneqq T^*(F^-|_U) \cong T^*U \times L^+G \times (K\fg/K^+\fg).
    \end{split}
\end{equation}
This implies that: 
\begin{itemize}
    \item The section \(TU \to TF^-|_U\) under the first identification is given by \(\sigma^{-1}d\sigma\). Vector bundles on \(F^-|_U\) which do not vary along \(L^+G\), i.e.\ regular maps \(a \colon F^-|_U \to TF^-|_U\) of the form
    \begin{equation}\label{eq:no_vary_LG}
        a(u,g) = (a_1(u),g,a_2(u))\,,\qquad u \in U, g\in L^+G
    \end{equation}
    for a section \(a_1 \colon U \to TU\) and a map \(U \to L^+\fg\),
    can be identified with 
    \begin{equation}\label{eq:trivialized_tangent}
        L^+\fg(U) \oplus \textnormal{Im}(\sigma^{-1}d\sigma);
    \end{equation}
    \item The elements of \(K^-_\sigma\fg(U)\) are precisely identified with one-forms on \(F^-|_U\), i.e.\ sections \(F^-|_U \to T^*F^-|_U\), which do not vary along \(L^+G\) in the sense that they are of a similar form as \eqref{eq:no_vary_LG}.
\end{itemize}

\subsubsection{Local coordinates on \(U\)}\label{sec:local_coordinates_on_U}
Let us assume that \(U\) is affine and that we may chose a coordinate system \(\{(u_\alpha,\partial_\alpha)\}_{\alpha = 1}^m\) of \(U\), where \(m = \dim(M_0) = (g-1)d\). This means that we have a set of independent functions \(\{u_\alpha\}_{\alpha = 1}^m \subseteq \Gamma(U,\sheafO_U)\) in the sense that \(\partial_\alpha = \partial/\partial u_\alpha \in \Gamma(U,TU)\) are well-defined derivations which from a \(\Gamma(U,\sheafO_U)\)-basis and satisfy \([\partial_\alpha,\partial_\beta]= 0\). It is always possible to make such choice; see \cite[Theorem A.5.1]{hotta_takeuchi_tanasaki_dmodules}.

For all \(\alpha \in \{1,\dots,m\}\), let us write \(\xi_\alpha \coloneqq  \sigma^{-1}\partial_\alpha \sigma \colon U \to L\fg\). Then \([\partial_\alpha,\partial_\beta] = 0\) implies\footnote{This follows from the fact that \(\partial_\alpha \xi_\beta - \partial_\beta \xi_\alpha +[\xi_\alpha,\xi_\beta]\) can be rewritten as
\[\partial_\alpha( \sigma^{-1}) \partial_\beta \sigma
        +\sigma^{-1}\partial_\alpha \partial_\beta \sigma -  \partial_\beta( \sigma^{-1}) \partial_\alpha \sigma
        -\sigma^{-1}\partial_\beta \partial_\alpha \sigma + \sigma^{-1}\partial_\alpha (\sigma)\sigma^{-1}\partial_\beta (\sigma)-\sigma^{-1}\partial_\beta (\sigma)\sigma^{-1}\partial_\alpha (\sigma) = 0,\]
        where \(0 = \partial_\alpha( \sigma^{-1} \sigma) = \partial_\alpha( \sigma^{-1})\sigma + \sigma^{-1} \partial_\alpha \sigma\).}
\begin{equation}\label{eq:derivative_of_xi}
    \partial_\alpha \xi_\beta - \partial_\beta \xi_\alpha +[\xi_\alpha,\xi_\beta] = 0
\end{equation} 
and \eqref{eq:trivialized_tangent} can be rewritten as
\begin{equation}\label{eq:direct_sum_decomposition}
    L\fg(U) = L^+\fg(U) \oplus \bigoplus_{\alpha = 1}^m\Gamma(U,\sheafO_U)\xi_\alpha \oplus L^-_\sigma\fg(U)
\end{equation}
using the symbols from Section \ref{sec:universal_sections_via_sigma}.

Observe that \(\Gamma(U,T^*U) \cong K^+\fg(U) \cap K^-_\sigma\fg(U)\) and we denote the image of \(du_\alpha\) under this isomorphism by \(\omega_\alpha\). It holds that
\begin{equation}\label{eq:omega_xi_dual}
    B(\xi_\alpha,\omega_\beta) = \delta_{\alpha\beta}
\end{equation}
using the bilinear form \(B\) from \eqref{eq:B}. In particular, 
\begin{equation}\label{eq:def_t}
    t = \sum_{\alpha=1}^m \omega_\alpha \otimes \xi_\alpha
\end{equation}
represents the canonical tensor of \(\Gamma(U,T^*U \otimes TU)\).  

\subsubsection{The derivations \(\nabla_\alpha\)}\label{sec:derivations}
Observe that \(L_\sigma^-\fg(U)\) and \(K_\sigma^-\fg(U)\) are stabilized by \(\nabla_\alpha \coloneqq \partial_\alpha +
\textnormal{ad}(\xi_\alpha)\), since it is straightforward\footnote{Indeed, \((\partial_\alpha + \textnormal{ad}(\xi_\alpha))(\sigma^{-1}a\sigma) = \partial_\alpha (\sigma^{-1}) a \sigma + \sigma^{-1} \partial_\alpha a \sigma + \sigma^{-1} a \partial_\alpha \sigma + \sigma^{-1}\partial_\alpha (\sigma) \sigma^{-1} a \sigma - \sigma^{-1} a \partial_\alpha \sigma = \sigma^{-1} \partial_\alpha a \sigma\) using \(\sigma^{-1}\partial_\alpha \sigma \sigma^{-1} = -\partial_\alpha (\sigma^{-1})\).} to calculate 
\((\partial_\alpha + \textnormal{ad}(\xi_\alpha))(\textnormal{Ad}(\sigma)^{-1}a) = \textnormal{Ad}(\sigma)^{-1} \partial_\alpha a\) holds for \(a \in L^-\fg(U)\) or \(a \in K^-\fg(U)\).

\subsubsection{Poisson structure on \(T^*F^-|_U\)}\label{sec:poisson_structure_restricted_to_U} Recall that vector bundles on \(F^-|_U\) which to not vary along \(L^+G\) are given by \(L^+\fg(U) \oplus \bigoplus_{\alpha = 1}^m\Gamma(U,\sheafO_U)\xi_\alpha\). In the following it will be useful to note that if we understand two such vector bundles \(a_1,a_2\) on \(F^-|_U\) as functions on
\[T^*F^-|_U \cong T^*U \times L^+G \times (K\fg/K^+\fg)\] 
constant along \(L^+G\), we have
\begin{equation}
    \{a_1,a_2\}(u) = \partial_{\pi(a_1(u))}a_2(u) - \partial_{\pi(a_2(u))}a_1(u) + [a_1(u),a_2(u)].
\end{equation}
Here, \(\pi \colon L\fg(U) \to L\fg(U)\) is the canonical projection onto \(\bigoplus_{\alpha = 1}^m\Gamma(U,\sheafO_U)\xi_\alpha\) with respect to the decomposition \eqref{eq:direct_sum_decomposition}.

This can be further rewritten as 
\begin{equation}\label{eq:poisson_structure_restricted_to_U}
    \{a_1,a_2\} =  [a_1,a_2] + \sum_{\alpha = 1}^m \left(B(\omega_\alpha,a_1)\partial_{\alpha}a_2 - B(\omega_\alpha,a_2)\partial_{\alpha}a_1\right)
\end{equation}
using \eqref{eq:omega_xi_dual} and the fact that \(\omega_\alpha \in K^+\fg \cap K_\sigma^-(U) = (L^+\fg(U) \oplus L^-_\sigma\fg(U))^\bot\).

\subsection{Construction of a dynamical \(r\)-matrix}\label{sec:rmatrix}
In the following, the \(\Gamma(U,\sheafO_U)\)-linear expansion of the bilinearform \(B\) introduced in \eqref{eq:B} will be denoted again by \(B\), so we have
\begin{equation}\label{eq:B_extension}
    B \colon L\fg(U) \times K\fg(U)\longrightarrow \Gamma(U,\sheafO_U).
\end{equation}
The projection \(\Pi_+ \colon L\fg(U) \to L\fg(U)\) onto \(L^+\fg(U)\) complementary to
\begin{equation}\label{eq:def_V}
    V \coloneqq \bigoplus_{\alpha = 1}^m\Gamma(U,\sheafO_U)\xi_\alpha \oplus L^-_\sigma\fg(U) \subseteq L\fg,    
\end{equation}
can be represented by a map \(r \colon U \to K\fg \widehat{\otimes} L^+\fg\) via
\begin{equation}\label{eq:r_matrix_projection_relation}
    B(a(u) \otimes 1, r(u)) = \Pi_+(a)(u)\,,\qquad a \in L\fg(U).
\end{equation}

\subsubsection{Coordinate expression of \(r\)}\label{sec:coordinate_r} Let \(z = (z_1,\dots,z_\ell)\) be a local coordinate of \(D\); see Section \ref{sec:coordinates}. Then \(K\fg \widehat{\otimes} L^+\fg \cong \prod_{i,j = 1}^\ell(\fg \otimes \fg)(\!(x_i)\!)[\![y_j]\!]dx_i\) and \(r\) has the form
\begin{equation}\label{eq:coordinate_expression_of_r}
    r(u;x,y) = \sum_{i = 1}^\ell\frac{\gamma dx_i}{x_i-y_i} + s(u;x,y)
\end{equation}
for any \(u\in U\). Here, \(s \colon U \to \prod_{i,j = 1}^\ell(\fg \otimes \fg)[\![x_i,y_j]\!] \) is some map, \(\gamma \coloneqq \sum_{i = 1}^{d} I_\alpha \otimes I_\alpha \in \fg \otimes \fg\) for a basis \(\{I_\alpha\}_{\alpha = 1}^{d}\subseteq \fg\) orthonormal with respect to \(\kappa\), and
\begin{equation}
    \frac{1}{x_i-y_i} = \sum_{k = 0}^\infty x_i^{-k-1}y_i^k\in \bC(\!(x_i)\!)[\![y_i]\!]
\end{equation}
for all \(i \in \{1,\dots,\ell\}\).
Indeed, \(s\) in \eqref{eq:coordinate_expression_of_r} is uniquely defined by the fact that \(r(u)\) is a generating series for \(V(u)^\bot \subseteq K\fg\), where
\begin{equation}\label{eq:V_u}
     V(u) \coloneqq \bigoplus_{\alpha = 1}^m \bC \xi_\alpha(u) \oplus \textnormal{Ad}(\sigma(u)^{-1})L^-\fg \subseteq L\fg.
\end{equation}
More precisely, \(r\) is the unique series of the form \eqref{eq:coordinate_expression_of_r} with the property 
\begin{equation}
    r(u) \in \prod_{i = 1}^\ell (V(u)^\bot \otimes \fg)[\![y_i]\!]. 
\end{equation}
Equivalently, we have
\begin{equation}\label{eq:V_u_bot}
     V(u)^\bot = \textnormal{Span}\{I_\alpha z_i^{-k-1} + s_{i,k,\alpha}(u;z)\mid 1\le i \le \ell,k \in \bN_0^\ell,1\le \alpha \le d\}
\end{equation}
where \(s\) was expanded as
\begin{equation}
    s(u;x,y) = \sum_{i = 1}^\ell \sum_{k = 0}^\infty \sum_{\alpha = 1}^{d} s_{i,k,\alpha}(u;x) \otimes I_\alpha y_i^{k}.
\end{equation}
Another equivalent perspective is that 
\begin{equation}
    r(u;x,y) =  \sum_{i = 1}^\ell \sum_{k = 0}^\infty \sum_{\alpha = 1}^{d} r_{i,k,\alpha}(u;x) \otimes I_\alpha y_i^{k}
\end{equation}
for the unique elements \(r_{i,k,\alpha}(u;z) = I_\alpha z_i^{-k-1} + s_{i,k,\alpha}(u;z)\) with the properties
\begin{equation}
    r_{i,k,\alpha}(u;z) \in V(u)^\bot \textnormal{ and }B(I_{\alpha_1} z_{i_1}^{k_1},r_{i_2,k_2,\alpha_2}(u;z)) = \delta_{i_1,i_2}\delta_{k_1,k_2}\delta_{\alpha_1,\alpha_2}.
\end{equation}
Observe that this interpretation implies that \((\textnormal{Ad}(\sigma) \otimes 1)r\) takes values in \(K^-\fg \widehat{\otimes} L^+\fg\).

\subsubsection{Other projections in terms of \(r\)}
Let us write \(\Pi_- \coloneqq 1-\Pi_+ \colon L\fg(U) \to L\fg(U)\) for the projection onto \(V\) complementary to \(L^+\fg(U)\). The adjoint \(\Pi_+^*\colon K\fg(U) \to K\fg(U)\) of \(\Pi_+\) with respect to \(B\) is the projection onto \(V^\bot\) complementary to \(K^+\fg(U)\). Similarly, \(\Pi_-^* = 1 - \Pi_+^* \colon K\fg(U) \to K\fg(U)\) is the projection onto \(K^+\fg(U)\) complementary to \(V^\bot\).

These projections can all be expressed using \(r\) in a similar fashion as \eqref{eq:r_matrix_projection_relation}. Namely, we have 
\begin{equation}
    \Pi_-(a) = B(1 \otimes a,\overline{r}), 
\end{equation}
for all \(a \in L\fg\). Here, \(\overline{r} \colon U \to L\fg \widehat{\otimes} K^+\fg\) is defined in local coordinates via
\begin{equation}\label{eq:coordinate_expression_of_overliner}
    \overline{r}(u;x,y) = \sum_{i = 1}^\ell\frac{\gamma dy_i}{x_i-y_i} - \tau(s(u;y,x))
\end{equation}
for the \(\prod_{i,j = 1}^\ell \bC[\![x_i,x_j]\!]\)-linear extension \(\tau\) of the tensor factor switch \(a\otimes b\to b \otimes a\) of \(\fg \otimes \fg\).

Moreover, we have   
\begin{equation}
    \Pi^*_+(a) = B(1 \otimes a,r) \textnormal{ and }\Pi^*_-(a) = B(a \otimes 1,\overline{r})
\end{equation}
for \(a \in K\fg\).

\subsubsection{Algebraicity of \(r\) in \(U\)}
Observe that \(K^+\fg \widehat{\otimes} L^+\fg \cong \prod_{i,j = 1}^\ell(\fg \otimes \fg)[\![x_i,y_j]\!]\) is an affine scheme of infinite type in a similar fashion as \(K^+\fg \cong \prod_{i = 1}^\ell\fg[\![z_i]\!]dz_i\) and \(L^+\fg \cong \prod_{i = 1}^\ell\fg[\![z_i]\!]\) are.

The map \(r \colon U \to K\fg \widehat{\otimes} L^+\fg\) is regular in the sense that \(s \colon U \to K^+\fg \widehat{\otimes} L^+\fg\) defined by \eqref{eq:coordinate_expression_of_r} is regular. Indeed, this follows from the fact that \(\Pi_+\) is uniquely defined by a regular map 
\begin{equation}
    U \longrightarrow \textnormal{Hom}(L\fg/L^+\fg,L^+\fg),    
\end{equation}
and the latter space can be identified with \(K^+\fg \widehat{\otimes} L^+\fg\) via \(B\) as affine schemes. The image under this identification is precisely \(s\).
 
\subsection{Theorem (Dynamical classical Yang-Baxter equation for \(r\))}\label{thm:dynamicalCYBE}
    Using the notation of sections \ref{sec:trivialization_ofF-} and \ref{sec:rmatrix}, the tensor \(r\) satisfies the following dynamical version of the classical Yang-Baxter equation
    \begin{equation}\label{eq:dynamicalCYBE}
    \begin{split}&[\overline{r}^{(12)},r^{(13)}] + [r^{(12)},r^{(23)}] + [r^{(13)},r^{(23)}] = \sum_{\alpha = 1}^m \left(\omega_\alpha^{(1)}\partial_\alpha r^{(23)} - \omega_\alpha^{(2)}\partial_\alpha r^{(13)}\right).
    \end{split}
    \end{equation}
    Here, under consideration of \(\fg \subseteq \textnormal{Mat}_{n\times n}(\bC)\), the notations \((\cdot)^{(i)},(\cdot)^{ij}\) can be understood coefficient-wise as e.g.\ \(a^{(2)} = 1 \otimes a \otimes 1, (a\otimes b)^{13} = a \otimes 1 \otimes b \in \textnormal{Mat}_{n\times n}(\bC)^{\otimes 3}\) and the Lie brackets are understood as coefficient-wise commutators in \(\textnormal{Mat}_{n\times n}(\bC)^{\otimes 3}\).
\subsection{Proof of Theorem \ref{thm:dynamicalCYBE}} 
The proof proceeds by identifying both sides of  \eqref{eq:dynamicalCYBE} with the failure of \(V = \bigoplus_{\alpha = 1}^m\Gamma(U,\sheafO_U)\xi_\alpha \oplus L^-_\sigma\fg(U)\) to be a subalgebra. 

\subsubsection{Identifying the left-hand side of \eqref{eq:dynamicalCYBE}} Consider the function \(\phi \colon U \to K^+\fg \widehat{\otimes} K^+\fg \widehat{\otimes} L^+\fg\) uniquely determined by the property
    \begin{equation}\label{eq:definition_phi}
        B(a \otimes b \otimes c, \phi) = B([a,b],c)\,,\qquad a,b \in V, c \in V^\bot. 
    \end{equation}
    If \(\ell = 1\), so \(S = \{p\}\), the left-hand side of \eqref{eq:dynamicalCYBE} is equal to \(\phi\) at any \(u \in U\) by virtue of \cite[Theorem 3.6]{abedin_maximov_stolin_quasibialgebras} under consideration of \eqref{eq:coordinate_expression_of_r}. The proof of \cite[Theorem 3.6]{abedin_maximov_stolin_quasibialgebras} can be easily adjusted to see that the left-hand side of \eqref{eq:dynamicalCYBE} remains equal to \(\phi\) for \(\ell > 1\), i.e.\ for general finite subsets \(S \subseteq X\).

    \subsubsection{Identifying the right-hand side of \eqref{eq:dynamicalCYBE}}
    Let us write 
    \begin{equation}
        \psi \coloneqq \sum_{\alpha = 1}^m \left(\omega_\alpha^{(1)}\partial_\alpha r^{(23)} - \omega_\alpha^{(2)}\partial_\alpha r^{(13)}\right) \colon U \longrightarrow K^+\fg \widehat{\otimes} K^+\fg \widehat{\otimes} L^+\fg.
    \end{equation}
    It remains to prove that \(\psi = \phi\), or equivalently, 
    \begin{equation}\label{eq:definition_psi}
        B(a \otimes b \otimes c, \psi) = B([a,b],c)\,,\qquad a,b \in V, c \in V^\bot. 
    \end{equation}
    We have \([L^-_\sigma\fg(U),L^-_\sigma\fg(U)] \subseteq L^-_\sigma\fg(U)\) and \(V^\bot \subseteq L^-_\sigma \fg(U)^\bot = K^-_\sigma(U)\), so both sides of \eqref{eq:definition_psi} vanish for \(a, b\in L^-_\sigma\fg(U)\). Using the (skew-)symmetry as well as the \(\Gamma(U,\sheafO_U)\)-linearity of \(B\) and the Lie bracket, it remains to verify \eqref{eq:definition_phi} in the following two cases:
    \begin{enumerate}
        \item \(a = \xi_\alpha\) and \(b = \textnormal{Ad}(\sigma)^{-1}\widetilde{b}\) for \(\widetilde{b} \in L^-\fg\);

        \item \(a = \xi_\alpha\) and \(b = \xi_\beta\).
    \end{enumerate}
    In Case (1) we have
    \begin{align}\label{eq:psi_phi_1}
         &B\left(\xi_\alpha \otimes b \otimes c,\psi\right) = B(b \otimes c, \partial_\alpha r) = - B(\partial_\alpha b \otimes c,r) = B([\xi_\alpha,b],c),
    \end{align}
    so \eqref{eq:definition_psi} is satisfied.
    Here, we used in the second equality that \(B(f \otimes 1,r) = 0\) for all \(f \in L^-_\sigma\fg(U)\) implies
    \begin{equation}\label{eq:derivative_of_B}
        0 = \partial_\alpha B(b \otimes c,r) = B(\partial_\alpha b \otimes c,r) + B(b \otimes \partial_\alpha c,r) + B(b \otimes c,\partial_\alpha r) = B(\partial_\alpha b \otimes c,r) + B(b \otimes c,\partial_\alpha r).
    \end{equation}
    Furthermore, in the last equality of \eqref{eq:psi_phi_1} we used that \(\nabla_\alpha b = \textnormal{Ad}(\sigma)^{-1}\partial_\alpha \widetilde{b} = 0\).
    
    In Case (2), we have 
    \begin{align*}
         &B\left(\xi_\alpha \otimes \xi_\beta \otimes c,\psi\right)  = B(\xi_\beta \otimes c, \partial_\alpha r) - B(\xi_\alpha \otimes c,\partial_\beta r) \\&= - B((\partial_\alpha \xi_\beta - \partial_\beta \xi_\alpha)  \otimes c,r) = B([\xi_\alpha,\xi_\beta],c).
    \end{align*}
    Here, similar arguments as in \eqref{eq:derivative_of_B} were used in the second equality and \(\partial_\alpha \xi_\beta - \partial_\beta \xi_\alpha +[\xi_\alpha,\xi_\beta] = 0\) implied the last equality.
    This concludes the proof.\hfill \boxed{}

    \subsection{Connection to the formal Hitchin system} First, we want to show that \(r\) is an \(r\)-matrix for the formal Hitchin system on \(T^*F^-\). To do so, we have to find a Lax representation of this system.
    
    \subsubsection{Lax representation} 
    Consider the canonical fiber-wise embedding \(L \colon T^*F^- \to K\fg\) and observe that the Hamiltonians of the formal Hitchin system is given by
    \begin{equation}
        \phi(L) \colon T^*F^- \longrightarrow Z,
    \end{equation}
    where \(\phi \in \textnormal{Sym}(\fg)^G\) and \(Z\) is introduced in Section \ref{sec:hitchin_fibration}.

    In order to show that \(L\) will define the Lax matrix of this system, we need the following lemma.

    \subsubsection{Lemma}\label{lem:poisson_and_r_bracket} Fix \(u_0 \in U\) and consider \(\theta \colon U \to \textnormal{Aut}_{\bC\textnormal{-alg}}(L\fg)\) defined by 
    \begin{equation}
        \theta(u) \coloneqq \textnormal{Ad}(\sigma(u)^{-1})\textnormal{Ad}(\sigma(u_0)).    
    \end{equation}
    Then \(\theta(u)\) defines an isomorphism from \(T_{[\sigma(u_0)]}F^-\) to \(T_{[\sigma(u)]}F^-\) for all \(u \in U\), so \(u\mapsto \theta(u)a\) defines a vector field \(\theta(a) \in L^+\fg(U) \oplus \bigoplus_{\alpha = 1}^m \Gamma(U,\sheafO_U)\xi_\alpha\) on \(F^-|_U\) for every \(a \in T_{[\sigma(u_0)]}F^-\) and we have
    \begin{equation}
        \{\theta(a),\theta(b)\} = [\Pi_+\theta(a),\Pi_+\theta(b)]-[\Pi_-\theta(a),\Pi_-\theta(b)] = \frac{1}{2}([R\theta(a),\theta(b)]-[\theta(a),R\theta(b)]).
    \end{equation}
    for \(R \coloneqq \Pi_+ - \Pi_-\) and all \(a \in T_{[\sigma(u_0)]}F^-\). Here, the projections \(\Pi_\pm\) were introduced in Section \ref{sec:rmatrix}.
    
    \subsubsection{Proof of Lemma \ref{lem:poisson_and_r_bracket}}
    The Poisson bracket reads 
    \begin{equation}
        \{\theta(a),\theta(b)\} = [\theta(a),\theta(b)] + \sum_{\alpha = 1}^m\left( B(\omega_\alpha,\theta(a))\partial_\alpha \theta(b) - B(\omega_\alpha,\theta(b))\partial_\alpha \theta(a)\right);
    \end{equation}
    see \eqref{eq:poisson_structure_restricted_to_U}.
    Using the fact that \(\nabla_\alpha \theta(v) = 0\) for all \(v \in T_{[\sigma(u_0)]}F^-\) implies \(\partial_\alpha \theta(v) = -[\xi_\alpha,\theta(v)]\) and that \(\Pi_-(v) = \sum_{\alpha = 1}^m B(\omega_\alpha,v)\xi_\alpha\) holds for all \(v \in L^+\fg(U) \oplus \bigoplus_{\alpha = 1}^m\Gamma(U,\sheafO_U)\xi_\alpha\), we obtain
    \begin{equation}
        \begin{split}
            \{\theta(a),\theta(b)\} &= [\theta(a),\theta(b)] - [\Pi_-\theta(a),\theta(b)] - [\theta(a),\Pi_-\theta(b)] \\&= [\Pi_+\theta(a),\Pi_+\theta(b)] - [\Pi_-\theta(a),\Pi_-\theta(b)].
        \end{split}
    \end{equation}

    \subsubsection{Local representation of \(L\)}
    Fix a point \(u_0 \in U\) and write \(\rho_0 = r(u_0) + \sum_{\alpha = 1}^m \omega_\alpha(u_0) \otimes \xi_\alpha(u_0)\) as well as \(\theta(u) \coloneqq \textnormal{Ad}(\sigma(u)^{-1})\textnormal{Ad}(\sigma(u_0))\). 
    Then we can determine \(L\) from the tensor \((\theta \otimes \theta)\rho_0\) in the sense that
    \begin{equation}\label{eq:lax_theta}
        v = L(v) = B(1 \otimes v,(\theta(u) \otimes \theta(u))\rho_0)
    \end{equation}
    for every \(u \in U\) and \(v \in T^*_uF^-\).

    \subsubsection{Proposition}\label{prop:lax_rep} Let \(H\) be a Hamiltonian of the Hitchin system and let 
    \begin{equation}
        Q \colon T^*F^-|_U \to L\fg\,,\qquad Q = \frac{1}{2} RdH(L).
    \end{equation}
    Then 
    \begin{equation}
        \frac{dL}{dt} \coloneqq \{H,L\} = [Q,L]
    \end{equation}
    holds, so \((L,Q)\) gives a Lax pair for the Hitchin system.

    \subsubsection{Proof of Proposition \ref{prop:lax_rep}}
    Lemma \ref{lem:poisson_and_r_bracket} and the fact that \(L\) is represented by a tensor of the form \((\theta \otimes \theta)\rho_0\) implies that
    \begin{equation}
        \{H,L\} = \frac{1}{2}([RdH(L),L] + [dH(L),RL]) = [Q,L]
    \end{equation}
    for \(Q = \frac{1}{2}RdH(L)\). Here, we used that \([dH(v),v] = 0\) for all \(v \in K\fg\) since \(H \in \overline{\textnormal{Sym}}(L\fg)^{L\fg}\).

    \subsection{Theorem (The \(r\)-matrix of the formal Hitchin system)}\label{thm:rmatrix_of_Hitchin}
    The dynamical \(r\)-matrix \(r\) is an \(r\)-matrix of the formal Hitchin system
    \begin{equation}\label{eq:russion_formula}
        \{L \otimes L\}= [1 \otimes L,r] + [L \otimes 1,\overline{r}],
    \end{equation}
    where \(\{L \otimes L\}\colon U \to K\fg \otimes K\fg \) is defined by
    \begin{equation}
        B(a(u) \otimes b(u),\{L\otimes L\}(c)) = B(\{a,b\}(u),c)
    \end{equation}
    for all \(u \in U\), \(a ,b \in L^+\fg(U) \oplus \bigoplus_{\alpha = 1}^m \Gamma(U,\sheafO_U)\xi_\alpha\), and \(c \in T^*_{[\sigma(u)]}F^-\).
    
    Let us note here that replacing \(\overline{r}\) with the more commonly used notation \(-r^{(21)}\) gives \eqref{eq:intro_Lax}.
    \subsubsection{Proof of Theorem \ref{thm:rmatrix_of_Hitchin}}
    The equation \eqref{eq:russion_formula} is equivalent, by definition and \eqref{eq:lax_theta}, to the fact that
    \begin{equation}\label{eq:calculation_rmatrix_hitchin}
        B(\{\theta(a),\theta(b)\},c) = B(\theta(a) \otimes \theta(b),[1 \otimes c,r] + [c \otimes 1,\overline{r}])
    \end{equation}
    holds for all \(a,b \in T_{[\sigma(u_0)]}F^-\) and \(c \in K^-_\sigma(U)\).
    
    Using 
    \begin{equation}
        B(\xi_\alpha \otimes 1,r) = 0 = B(1 \otimes a,\overline{r}) \textnormal{ and }B(1 \otimes \xi_\alpha,\overline{r}) = \xi_\alpha
    \end{equation}
    for all \(\alpha \in \{1,\dots, m\}\) and \(a \in L^+\fg\) we can see that
    \begin{equation}
        B(v \otimes w, [1 \otimes c,r] + [c \otimes 1,\overline{r}]) = \begin{cases}
            B(w,[c,v]) = B([v,w],c) & v,w \in L^+\fg\\
            0 &v = \xi_\alpha, w \in L^+\fg\\
            B(\xi_\alpha,[c,\xi_\beta]) = -B([\xi_\alpha,\xi_\beta],c)& v = \xi_\alpha,w = \xi_\beta.
        \end{cases}
    \end{equation}
    holds. This implies
    \begin{equation}
        B(v \otimes w,[1 \otimes c,r] + [c \otimes 1,\overline{r}]) = B(c,[\Pi_+v,\Pi_+w]-[\Pi_-v,\Pi_-w])
    \end{equation}
    for all \(v,w \in L^+\fg(U) \oplus \bigoplus_{\alpha = 1}^m\Gamma(U,\sheafO_U)\xi_\alpha\).
    Therefore, Lemma \ref{lem:poisson_and_r_bracket} concludes the proof.

    \subsubsection{Specialization to the punctured Hitchin system on \(M_{0,S}\)}
    The trivializations 
    \[F^-|_U \cong U \times L^+G \textnormal{ and } M_{0,S}|_U \cong U \times G^\ell\]
    combined with the embedding \(G^\ell \subseteq L^+G\) induce a Poisson embedding \(T^*M_{0,S}|_U \subseteq T^*F^-|_U\). Therefore, over the \'etale open subset \(U\), the Hamiltonians and the Lax presentation the punctured Hitchin system is simply obtained by restriction from the those of the formal Hitchin system. In particular, we have: 

    \subsection{Corollary (The \(r\)-matrix structure of the punctured Hitchin system)}\label{cor:rmatrix_of_Hitchin} The restriction of \((L,Q)\) to \(T^*M_{0,S}|_U\) defines a Lax presentation for the punctured Hitchin system and \(r\) is an \(r\)-matrix of the punctured Hitchin system with respect to this Lax presentation. 
    
    \subsection{Quantization}\label{sec:quantization}
    The Lax matrix restricted to \(T^*M_{0,S}|_U\) can be written as 
    \[L(u) = \sum_{\alpha = 1}^m \omega_\alpha(u;x) \otimes \xi_\alpha(u;y) + \sum_{i = 1}^\ell r(u;x,p_i).\]
    Indeed, this follows from the fact that
    \begin{equation}
        \begin{split}
            T^*_{[\sigma(u)]}M_{0,S} &\cong \Gamma(X,\textnormal{Ad}(P(u))\otimes \Omega_X(S)) \cong K^-_\sigma(U) \cap \mathfrak{m}^{-1}K^+\fg \\&= \bigoplus_{\alpha = 1}^m\bC\omega_\alpha(u) \oplus \bigoplus_{i = 1}^\ell\bigoplus_{\alpha = 1}^d \bC r_{i,0,\alpha}(u) \subseteq K\fg.
        \end{split}
    \end{equation}
    and \(r(u;x,p_i) = \sum_{\alpha = 1}^d r_{i,0,\alpha}(u;x) \otimes I_{\alpha}^{(i)}\), where \([\sigma(u)]\) is the point in \(M_{0,S}\) defined by \(\sigma(u) \in LG\) and \(I_\alpha^{(i)} \in \fg^{\ell} \subseteq L\fg\) is the image of \(I_\alpha \in \fg\) in the \(i\)-th direct summand.
    
    The associated quadratic Hamiltonian \(H = \frac{1}{2}\kappa(L,L) \colon T^*M_{0,S}|_U \to \Gamma(X,\Omega_X(S)^{\otimes 2})\) reads
    \begin{equation*}
        \begin{split}
        H &=  \frac{1}{2}\sum_{\alpha,\beta = 1}^m \kappa(\omega_\alpha,\omega_\beta) \otimes \xi_\alpha\xi_\beta + \sum_{\alpha = 1}^m\sum_{i = 1}^\ell\sum_{\beta = 1}^d \kappa(\omega_\alpha, r_{i,0,\beta}) \otimes \xi_\alpha I_\beta^{(i)} \\&+ \frac{1}{2}\sum_{i = 1}^\ell\sum_{\alpha,\beta = 1}^d K(r_{i,0,\alpha},r_{j,0,\beta}) \otimes I^{(i)}_\alpha I^{(j)}_\beta 
        \end{split}
    \end{equation*}
    which is a map \(U \times G^\ell \to \Gamma(X,\Omega_X(S)^{\otimes 2}) \otimes \textnormal{Sym}(L\fg)\) 
    understood as a regular map 
    \[U \times G^\ell \times K\fg \to \Gamma(X,\Omega_X(S)^{\otimes 2})\]
    constant in \(G^\ell\) and restricted to \(T^*M_{0,S}\) via \(T^*M_{0,S}\subseteq U \times G^\ell \times K\fg\).

    We obtain an honest Hamiltonian associated to the puncture \(p_i\) by reading of the tensor coefficient of \(x_i^{-1}(dx_i)^2\) in local coordinates (see Section \ref{sec:coordinates}) in this expression. Using \eqref{eq:coordinate_expression_of_r}, this yields the Hamiltonian
    \begin{equation*}
        \begin{split}
        H_i(u) &= \sum_{\alpha = 1}^\ell \omega_\alpha(u;p_i)^{(i)}\xi_\alpha + s(u;p_i,p_i)^{(ii)}+\sum_{j = 1, j \neq i}^\ell r(u;p_i,p_j)^{(ij)} 
        \end{split}
    \end{equation*}
    which is understood as element of 
    \[\Gamma(T^*M_{0,S}|_U,\sheafO_{T^*M_{0,S}|_U}) \cong \Gamma(T^*U,\sheafO_{T^*U}) \otimes \Gamma(G^\ell,\sheafO_{G^\ell}) \otimes \textnormal{Sym}(\fg)^{\otimes \ell}\] 
    constant in \(G^\ell\), i.e.\ equals to 1 in the tensor factor \(\Gamma(G^\ell,\sheafO_{G^\ell})\). Here, \(\xi_\alpha \in \Gamma(U,TU)\) is understood as regular function on \(T^*U\) and \((a \otimes b)^{(ij)} \in \textnormal{Sym}(\fg)^{\otimes \ell}\) is given by \(a\) in the \(i\)-th tensor factor and \(b\) in the \(j\)-th tensor factor for all \(a,b\in \fg\). Furthermore, we identified \(\omega_\alpha(u;p_i)\) (resp.\ \(s(u;p_i,p_j), r(u;p_i,p_j)\)), which is an element of \(\fg dx_i\) (resp.\ which are elements of \((\fg \otimes \fg)dx_i\)), with its coefficient in \(\fg\) (resp. \(\fg \otimes \fg\)).

    In \cite{felder_kzb}, it is shown that 
    \begin{equation}
        H_i(u) = \sum_{\alpha = 1}^\ell \omega_\alpha(u;p_i)^{(i)}\partial_\alpha + s(u;p_i,p_i)^{(ii)}+\sum_{j = 1, j \neq i}^\ell r(u;p_i,p_j)^{(ij)} - h^\vee q(u;p_i)^{(i)}
    \end{equation}
    forms a set of commuting Hamiltonians, understood as differential operators on \(T^*M_{0,S}|_U\) with values in \(\textnormal{U}(\fg)^{\otimes \ell}\). This can be seen as a quantization in the standard way, i.e.\ replacing \(\xi_\alpha\) by \(\partial_\alpha\) and \(\textnormal{Sym}(\fg)\) by \(\textnormal{U}(\fg)\), up to the term \(h^\vee q(u;p_i)^{(i)}\) which is a quantum correction coming from central extension of \(L\fg\) and passing to the critical level. In fact, these Hamiltonians are obtained in the critical limit from the KZB connection on the bundle of conformal blocks, which can also be expressed using \(r\).
    
    \section{The extended \(r\)-matrix} \label{ch:ex_rmatrices}
    \subsection{Generalized Szeg\"o kernel of regularly stable \(G\)-bundles}\label{sec:algebro_geometric_construction}
    For \(Q \in M_0\) it is well known that \(\textnormal{H}^0(\textnormal{Ad}(Q)) = 0\). Therefore, \(\textnormal{H}^1(\textnormal{Ad}(Q) \otimes \Omega_X) = 0\) holds as well by Serre duality under consideration of \(\textnormal{Ad}(Q)^* \cong \textnormal{Ad}(Q)\). Consider the short exact sequence
    \begin{equation}\label{eq:residue_sequence}
        0   \longrightarrow (\textnormal{Ad}(Q) \otimes \Omega_X) \boxtimes \textnormal{Ad}(Q) \longrightarrow ((\textnormal{Ad}(Q) \otimes \Omega_X) \boxtimes \textnormal{Ad}(Q))(\Delta) \stackrel{\textnormal{res}_\Delta}\longrightarrow \delta_*(\textnormal{Ad}(Q) \otimes \textnormal{Ad}(Q)) \longrightarrow 0
    \end{equation}
    defined by taking the residue along the diagonal divisor \(\Delta = \textnormal{Im}(\delta)\), where \(\delta\colon X \to X \times X\) is given by \(p \mapsto (p,p)\); see e.g.\ \cite{burban_galinat} for a detailed definition.
    
    If we apply \(\Gamma(X \times X^\circ,-)\) to \eqref{eq:residue_sequence} under consideration of \(\textnormal{H}^0(\textnormal{Ad}(Q)) = 0 = \textnormal{H}^1(\textnormal{Ad}(Q) \otimes \Omega_X)\) we obtain
    \begin{equation}
        \begin{split}
        K \coloneqq \frac{ \textnormal{H}^0(((\textnormal{Ad}(Q) \otimes \Omega_X) \boxtimes \textnormal{Ad}(Q)|_{X^\circ})(\Delta))}{\textnormal{H}^0(\textnormal{Ad}(Q) \otimes \Omega_X) \otimes \Gamma(X^\circ,\textnormal{Ad}(Q)))}\stackrel{\cong}\longrightarrow \Gamma(X^\circ,\textnormal{Ad}(Q) \otimes \textnormal{Ad}(Q))
        \end{split}
    \end{equation}
    Here, the K\"unneth formulas \(\textnormal{H}^0((\textnormal{Ad}(Q) \otimes \Omega_X) \boxtimes \textnormal{Ad}(Q)|_{X^\circ}) = \textnormal{H}^0(\textnormal{Ad}(Q) \otimes \Omega_X) \otimes \Gamma(X^\circ,\textnormal{Ad}(Q))\) and
    \begin{equation}
        \begin{split}
        &\textnormal{H}^1((\textnormal{Ad}(Q) \otimes \Omega_X) \boxtimes \textnormal{Ad}(Q)|_{X^\circ}) \\&= (\textnormal{H}^1(\textnormal{Ad}(Q) \otimes \Omega_X) \otimes \Gamma(X^\circ,\textnormal{Ad}(Q))) \oplus (\textnormal{H}^0(\textnormal{Ad}(Q) \otimes \Omega_X) \otimes \textnormal{H}^1(\textnormal{Ad}(Q)|_{X^\circ})) \\&= 0
        \end{split}    
    \end{equation}
    were used.
    
    There is a unique element of \(\varrho \in K\) which is mapped to the identity of
    \begin{equation}
        \Gamma(X^\circ,\textnormal{Ad}(Q) \otimes \textnormal{Ad}(Q)) \cong \textnormal{End}_{\sheafO_{X^\circ}}(\textnormal{Ad}(Q)|_{X^\circ}).    
    \end{equation}
    This can be viewed as a generalization of the so-called Szeg\"o kernel in the sense of e.g.\ \cite{benzvi_biswas_szegokernels}. Moreover, \(\varrho\) should be seen as a higher genus analog of the algebro-geometric construction of solutions of the generalized classical Yang-Baxter equation from \cite{cherednik_definition_of_tau,burban_galinat}.  
    Indeed, in the following, we will relate \(\varrho\) to \(r\) and deduce an analog of the classical Yang-Baxter equation for \(\varrho\) to underline this point of view. Additionally, we will derive an analog of Theorem \ref{thm:rmatrix_of_Hitchin} for \(\varrho\).

    \subsubsection{Realization of \(\varrho\)}
    Consider 
    \begin{equation}
        \rho = r + t \colon U \to K\fg \widehat{\otimes} L\fg,    
    \end{equation}
    where \(t = \sum_{\alpha = 1}^m \omega_\alpha \otimes \xi_\alpha\). This is a reproducing kernel for the pairing between \(L\fg(U) \oplus \bigoplus_{\alpha = 1}^m\Gamma(U,\sheafO_U)\xi_\alpha\) and \(K_\sigma^-\fg(U)\), i.e.\
    \begin{equation}\label{eq:rho_and_B}
        B(a \otimes b,\rho) = B(a,b)
    \end{equation}
    holds for all \(a \in L\fg(U) \oplus \bigoplus_{\alpha = 1}^m\Gamma(U,\sheafO_U)\xi_\alpha\) and \(b \in K_\sigma^-\fg(U)\).
    
    \subsubsection{Proposition}\label{sec:globality_of_rho} The expression \(\rho\) is a point-wise representative of the generalized Szeg\"o kernel described in Section \ref{sec:algebro_geometric_construction}. In particular, \begin{equation}\label{eq:globality_of_rho}
        \rho(u) \in \textnormal{H}^0(((\textnormal{Ad}(P(u)) \otimes \Omega_X) \boxtimes \textnormal{Ad}(P(u))|_{X^\circ})(\Delta)), 
    \end{equation}
    where we recall that \(P(u)\) is the \(G\)-bundle associated to \(\sigma(u) \in LG\), and
    \begin{equation}\label{eq:residue_of_rho}\textnormal{res}_{\Delta}\rho(u) \in \Gamma(X^\circ, \textnormal{Ad}(P(u)) \otimes \textnormal{Ad}(P(u))) \cong \textnormal{End}_{\sheafO_{X^\circ}}(\textnormal{Ad}(P(u)))
    \end{equation}
    is the identity.
    \subsubsection{Proof of \eqref{sec:globality_of_rho}} Since \((\textnormal{Ad}(\sigma(u)) \otimes 1)r(u) \in K^-\fg \widehat{\otimes} L^+\fg\), we have 
    \begin{equation}
        (\textnormal{Ad}(\sigma(u)) \otimes 1)\rho(u) \in K^-\fg \widehat{\otimes} L\fg.    
    \end{equation}
    On the other hand, in the notation of Section \ref{sec:coordinate_r}, we can expand \(r(u)\) from \eqref{eq:coordinate_expression_of_r} in \(x\) to obtain
    \begin{equation}\label{eq:r_other_expansion}
        r = -\sum_{i = 1}^\ell\sum_{k = 0}^\infty\sum_{\alpha = 1}^{d} I_\alpha x_i^k dx_i \otimes \overline{r}_{i,k,\alpha}.
    \end{equation}
    Here, \(\overline{r}_{i,k,\alpha}\) is uniquely determined by
    \begin{equation*}
        \overline{r}_{i,k,\alpha}(u) \in V(u) = \bigoplus_{\alpha = 1}^m\bC \xi_\alpha(u) \oplus \textnormal{Ad}(\sigma(u)^{-1})L^-\fg \textnormal{ and }B(\overline{r}_{i_1,k_1,\alpha_1},I_\alpha x_i^k dx_i) = \delta_{i_1,i_2}\delta_{k_1,k_2}\delta_{\alpha_1,\alpha_2}.
    \end{equation*}
    Since \(B(\omega_\alpha,\xi_\beta) = \delta_{\alpha,\beta}\) for \(\alpha,\beta \in \{1,\dots,m\}\), we can rewrite \eqref{eq:r_other_expansion} as 
    \begin{equation}
        r = - \sum_{\alpha = 1}^m \omega_\alpha \otimes \xi_\alpha -\sum_{i = 1}^\ell\sum_{k = 0}^\infty\sum_{\alpha = 1}^m
        b_{i,k,\alpha} \otimes v_{i,k,\alpha},
    \end{equation}
    where for \(i \in \{1,\dots,\ell\},\alpha \in \{1,\dots,d\}\) and \(k \in \bN_0\) we took
    \begin{equation}
        b_{i,k,\alpha} = I_\alpha x_i^kdx_i  - \sum_{\beta = 1}^m B(\xi_\beta,I_\alpha x_i^kdx_i)\omega_\beta \textnormal{ and }v_{i,k,\alpha} = \overline{r}_{i,k,\alpha}  - \sum_{\beta = 1}^m B(\overline{r}_{i,k,\alpha},
        \omega_\alpha)\xi_\beta.
    \end{equation}
    By construction, \(B(v_{i,k,\alpha},\omega_\beta) = 0\), so \(v_{i,k,\alpha} \in \textnormal{Ad}(\sigma(u)^{-1})L^-\fg\). Therefore, \((1 \otimes \textnormal{Ad}(\sigma(u)))\rho(u)\) is actually an element of \(K^+\fg \widehat{\otimes} L^-\fg\).
    Gluing these two expressions together under consideration of
    \begin{equation}
        \Gamma(X^\circ,\textnormal{Ad}(P(u))) = \textnormal{Ad}(\sigma(u)^{-1})L^-\fg \textnormal{ and }\Gamma(X^\circ,\textnormal{Ad}(P(u)) \otimes \Omega_X) = \textnormal{Ad}(\sigma(u)^{-1})K^-\fg    
    \end{equation}
    results in the validity of \eqref{eq:globality_of_rho}.
    
     We can see that \(\textnormal{res}_{\Delta}\rho(u)\) is mapped to the identity in \eqref{eq:residue_of_rho} by restricting to \(D \times D\) and using  \eqref{eq:coordinate_expression_of_r} as well as the fact that \(\gamma\) is mapped to the identity under the isomorphism \(\fg \otimes \fg \cong \textnormal{End}(\fg)\) defined by \(\kappa\). \hfill \boxed{}

    \subsection{Theorem (Dynamical classical Yang-Baxter equation for \(\rho\))}\label{thm:extended_DCYBE}
    The expression \(\rho =  r + t\) satisfies
    \begin{equation}
    \begin{split}[\overline{\rho}^{(12)},\rho^{(13)}] + [\rho^{(12)},\rho^{(23)}] + [\rho^{(13)},\rho^{(23)}]  = \sum_{\alpha = 1}^m \left(\omega_\alpha^{(1)}\nabla_\alpha \rho^{(23)} - \omega_\alpha^{(2)}\nabla_\alpha \rho^{(13)}\right).
    \end{split}
    \end{equation}
    Here, \(\overline{\rho} = \overline{r} - \sum_{\alpha = 1}^m \xi_\alpha \otimes \omega_\alpha\) and the derivations \(\nabla_\alpha\) introduced in Section \ref{sec:derivations} act on tensors as \( \partial_\alpha + \textnormal{ad}(\xi_\alpha) \otimes 1 + 1\otimes \textnormal{ad}(\xi_\alpha)\).

    \subsection{Proof of Theorem \ref{thm:extended_DCYBE}}
    Writing \(t^{(21)} = \sum_{\alpha = 1}^m\xi_\alpha \otimes \omega_\alpha \otimes 1\), we can calculate:
    \begin{equation*}
        \begin{split}
            &[\overline{\rho}^{(12)},\rho^{(13)}] + [\rho^{(12)},\rho^{(23)}] + [\rho^{(13)},\rho^{(23)}]
            \\&= [\overline{r}^{(12)},r^{(13)}] + [r^{(12)},r^{(23)}] + [r^{(13)},r^{(23)}] + {[\overline{r}^{(12)},t^{(13)}]} + {
            [t^{(12)},r^{(23)}] + [t^{(13)},r^{(23)}]} \\&+ {[r^{(13)},t^{(21)}]} + {[r^{(12)},t^{(23)}]} + {[r^{(13)},t^{(23)}]} + {[t^{(13)},t^{(21)}] + [t^{(12)},t^{(23)}]} + {[t^{(13)},t^{(23)}]}
            \\&=
            [\overline{r}^{(12)},r^{(13)}] + [r^{(12)},r^{(23)}] + [r^{(13)},r^{(23)}] + {\color{red} [\overline{r}^{(12)},t^{(13)}]} + {\color{blue}
            [t^{(12)},r^{(23)}] + [t^{(13)},r^{(23)}]} \\&+ {\color{blue}[r^{(13)},t^{(21)}]} + {\color{red}[r^{(12)},t^{(23)}]} + {\color{blue}[r^{(13)},t^{(23)}]} + {\color{cyan}[t^{(13)},t^{(21)}] + [t^{(12)},t^{(23)}] + 2[t^{(13)},t^{(23)}]} {\color{red}-[t^{(13)},t^{(23)}]} \\& = \sum_{\alpha = 1}^m
            \left(\omega_\alpha^{(1)}\partial_\alpha r^{(23)} - \omega_\alpha^{(2)}\partial_\alpha r^{(13)} + {\color{red} \omega_\alpha^{(1)}\partial_\alpha t^{(23)} - \omega_\alpha^{(2)} \partial_\alpha t^{(13)}} \right.\\&  \left. + \,{\color{blue}\omega^{(1)}_\alpha [\xi_\alpha^{(2)} +\xi_\alpha^{(3)},r^{(23)}] - \omega_\alpha^{(2)} [\xi_\alpha^{(1)} +\xi_\alpha^{(3)},r^{(13)}]} + {\color{cyan} \omega^{(1)}_\alpha [\xi_\alpha^{(2)} +\xi_\alpha^{(3)},t^{(23)}] - \omega_\alpha^{(2)} [\xi_\alpha^{(1)} +\xi_\alpha^{(3)},t^{(13)}]}\right)  
            \\& = \sum_{\alpha = 1}^m \left(\omega_\alpha^{(1)}\nabla_\alpha \rho^{(23)} - \omega_\alpha^{(2)}\nabla_\alpha \rho^{(13)}\right).
        \end{split}
    \end{equation*}
    For the identification of the black terms in the last step, we used \eqref{eq:dynamicalCYBE}.
    For the identification of the {\color{red}red} terms, we used that 
    \begin{equation}\label{eq:vanishing_term_extended_DCYBE}
        [\overline{r}^{(12)},t^{(13)}] + [r^{(12)},t^{(23)}] = \sum_{\alpha,\beta = 1}^m \left(\omega_\alpha \otimes \partial_\alpha \omega_\beta \otimes \xi_\beta - \partial_\alpha \omega_\beta \otimes \omega_\alpha \otimes \xi_\beta\right)
    \end{equation}
    and \([\xi_\alpha,\xi_\beta]= -\partial_\alpha \xi_\beta + \partial_\beta \xi_\alpha\).
    It remains to prove \eqref{eq:vanishing_term_extended_DCYBE}.

    \subsubsection{Proof of \eqref{eq:vanishing_term_extended_DCYBE}}
    First of all, the equation is equivalent to proving 
    \begin{equation}
        [\overline{r},\omega_\alpha \otimes 1] + [r,1 \otimes \omega_\alpha] = \sum_{\beta = 1}^m \left(\omega_\beta \otimes \partial_\beta \omega_\alpha - \partial_\beta \omega_\alpha \otimes \omega_\beta\right).
    \end{equation}
    for all \(\alpha \in \{1,\dots,m\}\).
    Using \eqref{eq:coordinate_expression_of_r}, one can see that both sides of the equation are in \(K^+\fg \widehat{\otimes} K^+\fg\). Therefore, we have to show that for all \(a\in \{\xi_\alpha\}_{\alpha = 1}^m \cup L^-_\sigma\fg(U)\)
    \begin{equation}\label{eq:vanishing_original_equation}
        B(a \otimes b, [\overline{r},\omega_\alpha \otimes 1]+ [r,1 \otimes \omega_\alpha]) = \sum_{\beta}B(a \otimes b,\omega_\beta \otimes \partial_\beta \omega_\alpha - \partial_\beta \omega_\alpha \otimes \omega_\beta)
    \end{equation}
    holds. Observe that \(B(1 \otimes b,\overline{r}) = \Pi_-(b) = b\) and \(B(a \otimes 1, r) = \Pi_+(a) = 0\) implies
    \begin{equation}
        B(a \otimes b,[\overline{r},\omega_\alpha \otimes 1]+ [r,1 \otimes \omega_\alpha]) =B(a,[b,\omega_\alpha])=B([a,b],\omega_\alpha),
    \end{equation}
    so it remains to prove that
    \begin{equation}\label{eq:vanishing_term_in_bilinear_form}
        B([a,b],\omega_\alpha) = \sum_{\beta}B(a \otimes b,\omega_\beta \otimes \partial_\beta \omega_\alpha - \partial_\beta \omega_\alpha \otimes \omega_\beta).
    \end{equation}
    
    For \(a,b \in L^-_\sigma\fg(U)\) we have \(B(a,\omega_\alpha) = B(b,\omega_\beta) = B([a,b],\omega_\alpha) = 0\), so
    \begin{equation}
        \begin{split}
            \sum_{\beta = 1}^mB(a \otimes b,\omega_\beta \otimes \partial_\beta \omega_\alpha - \partial_\beta \omega_\alpha \otimes \omega_\beta) = 0 = B([a,b],\omega_\alpha)
        \end{split}
    \end{equation}
    and thus \eqref{eq:vanishing_term_in_bilinear_form} is satisfied.
    
    Next, we recall that \(b \in L_\sigma^-\fg(U)\) implies 
    \[\nabla_\gamma b = \partial_\gamma b + [\xi_\gamma,b] \in L_\sigma^-(U),\] 
    so \(B(\partial_\gamma b,\omega_\alpha) = -B([\xi_\gamma,b],\omega_\alpha)\). This and \(0 = \partial_\gamma B(b,\omega_\alpha) = B(\partial_\gamma b,\omega_\alpha) + B(b,\partial_\gamma \omega_\alpha)\) gives
    \begin{equation}
        \begin{split}
            \sum_{\beta = 1}^mB(\xi_\gamma \otimes b,\omega_\beta \otimes \partial_\beta \omega_\alpha - \partial_\beta \omega_\alpha \otimes \omega_\beta) = B(b,\partial_\gamma \omega_\alpha) = -B(\partial_\gamma b, \omega_\alpha) = B([\xi_\gamma,b],\omega_\alpha),
        \end{split}
    \end{equation}
    so \eqref{eq:vanishing_term_in_bilinear_form} is satisfied.

    Finally, consider
    \begin{equation}
        \begin{split}
            &\sum_{\beta}B(\xi_\gamma \otimes \xi_\delta,\omega_\beta \otimes \partial_\beta \omega_\alpha - \partial_\beta \omega_\alpha \otimes \omega_\beta) = B(\xi_\delta,\partial_\gamma \omega_\alpha) - B(\xi_\gamma,\partial_\delta\omega_\alpha) \\&=  -B(\partial_\gamma \xi_\delta - \partial_\delta \xi_\gamma,\omega_\alpha) = B([\xi_\gamma,\xi_\delta],\omega_\alpha)
        \end{split}
    \end{equation}
    concluding the proof.\hfill \boxed{}

    \subsection{Proposition (Analog of Theorem \ref{thm:rmatrix_of_Hitchin} for \(\rho\))}\label{thm:extended_rmatrix_of_hitchin} 
    The identity
    \begin{equation}\label{eq:extended_rmatrix_of_hitchin}
        B([a,b],c) = B(a \otimes b,[1 \otimes c,\rho]+[c \otimes 1,\overline{\rho}]) 
    \end{equation}
    holds for all \(a, b \in L^+\fg(U) \oplus \bigoplus_{\alpha = 1}^m\Gamma(U,\sheafO_U)\xi_\alpha\) and \(c \in K^-_\sigma\fg(U)\). 

    If we let \([L\otimes L] \colon T^*F^-|_U \to K\fg \otimes K\fg\) be the map uniquely determined by 
    \[B(a \otimes b, [L \otimes L](c)) = B([a,b],c)\]
    for all \(u \in U\),\(a, b \in T_{[\sigma(u)]}F^-\), and \(c \in T_{[\sigma(u)]}^*F^-\), \eqref{eq:extended_rmatrix_of_hitchin} can be rewritten as
    \begin{equation}
        [L \otimes L] = [1 \otimes L,\rho]+[L \otimes 1,\overline{\rho}].
    \end{equation}
    Writing \(\overline{\rho}\) as \(-\rho^{(21)}\), this takes the form \eqref{eq:intro_russian_rho}.
\subsection{Proof of Proposition \ref{thm:extended_rmatrix_of_hitchin}}
First of all, since \eqref{eq:extended_rmatrix_of_hitchin} is \(\Gamma(U,\sheafO_U)\)-linear in \(c\) on both sides, we may assume \(\textnormal{Ad}(\sigma)c \in K^-\fg\) is a constant function on \(U\). Recall that \(B(1 \otimes c,\rho) = c\) holds by virtue of \eqref{eq:rho_and_B}.
Therefore,
\begin{equation}\label{eq:rnatrix_of_Hitchin_using_DCYBE}
    \begin{split}
        &B(a\otimes b,[1 \otimes c,\rho] + [c \otimes 1,\overline{\rho}]) = B(a \otimes b \otimes c,[\rho^{(23)},\rho^{(12)}]+[\rho^{(13)},\overline{\rho}^{(12)}]) \\&= 
        B\left(a \otimes b \otimes c, [\rho^{(13)},\rho^{(23)}] - \sum_{\alpha = 1}^m \left(\omega_\alpha^{(1)}\nabla_\alpha \rho^{(23)} - \omega_\alpha^{(2)}\nabla_\alpha \rho^{(13)}\right)\right).
    \end{split}
\end{equation}
holds for all \(a,b \in L\fg(U)\) by virtue of Theorem \ref{thm:extended_DCYBE}.

Now \(B(v \otimes 1,\rho) =v\) for \(v \in L^+\fg(U) \oplus \bigoplus_{\alpha = 1}^m\Gamma(U,\sheafO_U)\xi_\alpha\), so
\begin{equation}\label{eq:commutator_in_r_poisson_bracket}
    B(a \otimes b \otimes c,[\rho^{(13)},\rho^{(23)}]) = B([a,b],c)    
\end{equation}
holds if additionally \(a, b \in L^+\fg(U) \oplus \bigoplus_{\alpha = 1}^m\Gamma(U,\sheafO_U)\xi_\alpha\). 

It remains to show that \(B(v \otimes c,\nabla_\alpha \rho) = 0\) for all \(v\in L\fg(U)\) if \(c \in \textnormal{Ad}(\sigma^{-1})K^-\fg\). Indeed, we can calculate
\begin{equation}
    \begin{split}
    &B(v \otimes c,\nabla_\alpha \rho) = \partial_\alpha B(v \otimes c,\rho) - B(\nabla_\alpha v \otimes c,\rho) - B(v \otimes \nabla_\alpha c, \rho) \\&= \partial_\alpha B(v,c) - B(\nabla_\alpha v, c) = -B([\xi_\alpha,v],c) - B(v,[\xi_\alpha,c]) = 0.
    \end{split}
\end{equation}
Here, \(\nabla_\alpha c = 0\) and 
\[\partial_\alpha B(v,c) = B(\partial_\alpha v, c) + B(v,\partial_\alpha c) = B(\partial_\alpha v, c) -B(v, [\xi_\alpha,c])\]
was used.
This concludes the proof. \hfill \boxed{}

\subsection{Choosing a different quasi-section \(\sigma\)}\label{sec:choosing_a_different_sigma} If the quasi-section \(\sigma\) is replaced by another quasi-section \(\sigma'\colon V\to LG\), we have \(\sigma' = g_-\sigma g_+\) for some regular morphisms \(g_\pm \colon V \times_{M_0} U \to L^\pm G\). Let \(\xi_\alpha', \omega_\alpha', t', r'\) and \(\rho'\) be the objects constructed in the previous two section using \(\sigma'\) instead of \(\sigma\). Assume that our coordinate system \(\{(u_\alpha,\partial_\alpha)\}_{\alpha = 1}^m\) on \(U\) and \(V\) coincide in \(V\times_{M_0} U\).

Then we have the following relations:
\begin{align*}
    &\xi_\alpha' = g_+^{-1}\sigma^{-1}g_-^{-1}\partial_\alpha(g_-\sigma g_+) = \textnormal{Ad}(g_+)^{-1}\left(\xi_\alpha + \textnormal{Ad}(\sigma)^{-1}\left(g_-^{-1}\partial_\alpha g_-\right)\right) + g_+^{-1}\partial_\alpha g_+;
    \\
    &\omega_\alpha' = \textnormal{Ad}(g_+)^{-1} \omega_\alpha;
    \\
    &t' = \left(\textnormal{Ad}(g_+)^{-1}\otimes \textnormal{Ad}(g_+)^{-1}\right)\left(t + \sum_{\alpha = 1}^m\omega_\alpha \otimes \textnormal{Ad}(\sigma)^{-1}\left(g_-^{-1}\partial_\alpha g_-\right)\right) + \sum_{\alpha = 1}^m \textnormal{Ad}(g_+)^{-1}\omega_\alpha \otimes g_+^{-1}\partial_\alpha g_+;
    \\
    &r' = (\textnormal{Ad}(g_+)^{-1}\otimes \textnormal{Ad}(g_+)^{-1})r - \sum_{\alpha = 1}^m \textnormal{Ad}(g_+)^{-1}\omega_\alpha \otimes g_+^{-1}\partial_\alpha g_+;
    \\ 
    &\rho' = \left(\textnormal{Ad}(g_+)^{-1}\otimes \textnormal{Ad}(g_+)^{-1}\right)\left(\rho - \sum_{\alpha = 1}^m \omega_\alpha \otimes \textnormal{Ad}(\sigma)^{-1}\left(g_-^{-1}\partial_\alpha g_-\right)\right).
\end{align*}

\appendix
\section{Notation}\label{sec:not}
\begin{itemize}
    \item \(\textnormal{Mat}_{n \times n}(R)\) is the space of \(n \times n\)-matrices with entries in a ring \(R\) and \(GL_n(R) \subseteq \textnormal{Mat}_{n \times n}(R)\) is the subgroup of invertible matrices. Moreover, \(GL_n = GL_n(\bC)\) is also used for the general complex linear algebraic group.
    \item \(G \subseteq GL_n\) is a semisimple complex algebraic group of dimension \(d\) defined by an ideal \(I \subseteq \Gamma(GL_n,\sheafO_{GL_n})\) with Lie algebra \(\fg\).

    \item \(X\) is a Riemann surface, \(S = \{p_1,\dots,p_\ell\} \subseteq X\), \(D\) is the formal neighbourhood of \(S\), \(X^\circ = X\setminus S\), and \(D^\circ = D\setminus S\).
    
    \item \(\sheafO_X\) is the sheaf of regular functions on \(X\), \(O^+ \coloneqq \widehat\sheafO_{X,S} = \prod_{i = 1}^\ell \widehat\sheafO_{X,p_i}\), \(O^- \coloneqq \Gamma(X^\circ,\sheafO_X)\), and \(O\) is the quotient field of \(O^+\). 

    \item \(\Omega_X\) is the sheaf of regular 1-forms on \(X\), \(\Omega^+ \coloneqq \widehat{\Omega}_{X,S}\), \(\Omega^- \coloneqq \Gamma(X^\circ,\Omega_X)\), and \(\Omega = O\Omega^+\). 
    
    \item \(LG\) (resp.\ \(L^\pm G\)) is the ind-affine group representing the functor that assigns to any \(\bC\)-algebra \(R\) the group \(\textnormal{Hom}_{\bC\textnormal{-alg}}\left(\Gamma(G,\sheafO_G), R \widehat{\otimes} O\right)\) (resp.\ \(\textnormal{Hom}_{\bC\textnormal{-alg}}\left(\Gamma(G,\sheafO_G), R \widehat{\otimes} O^\pm\right)\)). The corresponding Lie algebras are \(L^\star\fg = \fg\otimes O^\star\), where \(\star \in  \{\emptyset,+,-\}\).

    \item \(K^\star\fg = \fg\otimes \Omega^\star\), where \(\star \in  \{\emptyset,+,-\}\).
    
    \item \(\underline{M}\) is the stack of \(G\)-bundles on \(X\). It is identified with \(L^-G\setminus LG \,/\, L^+G\). The substack of regularly stable \(G\)-bundles \(\underline{M}_0 \subseteq \underline{M}\) has a coarse moduli variety denoted by \(M_0\).

    \item For an \'etale quasi-section \(\sigma \colon U \to LG\) of \(LG_0 \to M_0\), we write \(L^-_\sigma\fg(U) \coloneqq \textnormal{Ad}(\sigma)L^-\fg(U)\) and \(K^-_\sigma\fg(U) \coloneqq \textnormal{Ad}(\sigma)K^-\fg(U)\).

    \item \(\pi\) is the projection onto \(\bigoplus_{\alpha = 1}^m\Gamma(U,\sheafO_U)\xi_\alpha\) with respect to the decomposition \eqref{eq:direct_sum_decomposition}.

    \item \(\Pi_+\) and \(\Pi_-\) are the projections onto \(L^+\fg(U)\) and \(V \coloneqq \bigoplus_{\alpha = 1}^m\Gamma(U,\sheafO_U)\xi_\alpha \oplus L^-_\sigma L(U)\) with respect to \eqref{eq:direct_sum_decomposition} respectively. Furthermore, \(\Pi_+^*\) and \(\Pi_-^*\) are the adjoint projections onto \(V^\bot\) and \(K^+\fg(U)\) with respect to \(K\fg(U) = K^+\fg(U) \oplus V^\bot\) respectively. 

    \item For a diagram \(S_1 \to S \leftarrow S_2\) of ind-schemes, \(S_1 \times_S S_2\) is the fiber product in the category of schemes and we drop the subscript \(S\) for \(S = \textnormal{Spec}(\bC)\).

    \item Using \(F^-_0 \to M_0 \stackrel{\sigma}\leftarrow U\) and \(M_{0,S} \to M_0 \stackrel{\sigma}\leftarrow U\), we write 
    \[F^-|_U = U \times_{M_0} F^-_0 \textnormal{ and }M_{0,S}|_U = U \times_{M_0} M_{0,S}\] 
    and these are identified with \(U \times L^+G\) and \(U\times G^\ell\) respectively. Furthermore, \(TF^-|_U \coloneqq T(F^-|_U)\) is identified with the fiber product of \(TF^- \to F^- \leftarrow F^-|_U\) and \(TU \times L^+G \times L^+\fg\) and we deal similarly with \(TF^-|_U,T^*M_{0,S}|U\), and \(TM_{0,S}|_U\). 
\end{itemize}

\printbibliography

\end{document}